\documentclass{article}

\usepackage[utf8]{inputenc}
\usepackage{amsmath}
\usepackage{mathtools}
\usepackage{amsfonts}
\usepackage{physics}
\usepackage{float}
\usepackage{graphicx}
\usepackage{verbatim}
\usepackage{epstopdf}

\usepackage{bbm}
\usepackage{bm}
\usepackage{algorithmic}
\usepackage{algorithm}
\usepackage[round]{natbib}
\usepackage{amsthm} 
\usepackage{algorithmic}
\usepackage{algorithm}
\usepackage{geometry}
\usepackage{mleftright }
\usepackage{tabularx}
    \newcolumntype{L}{>{\raggedright\arraybackslash}X}
\usepackage{epstopdf}
\usepackage[caption=false]{subfig}

\usepackage[round]{natbib}

\usepackage{amscd,amsfonts,amsmath,amssymb,amsthm,bbm,bm,latexsym,mathrsfs}
\usepackage{epsfig,graphics,graphicx,longtable}
\usepackage[english]{babel}
\usepackage[usenames]{color}
\usepackage{bookmark}
\usepackage{sgame}
\usepackage{gensymb}
\usepackage{epstopdf}
\usepackage{float}
\usepackage{comment}

\usepackage{tikz} 
\usetikzlibrary{matrix}
\usetikzlibrary{bayesnet}

\usepackage{geometry}

\makeatletter
\newcommand{\svast}{\bBigg@{3}}
\newcommand{\vast}{\bBigg@{4}}
\newcommand{\Vast}{\bBigg@{5}}
\makeatother

\begin{document}


\title{\bfseries \sffamily Bayesian Multivariate Nonlinear State Space Copula Models}
	\date{\small \today}
			\author{Alexander Kreuzer\footnote{Corresponding author: {E-mail: a.kreuzer@tum.de}} \footnotemark[3], Luciana Dalla Valle\footnotemark[2], and Claudia Czado\footnotemark[3]}
\date{%
	Technische Universit\"at M\"unchen\footnotemark[3] and University of Plymouth\footnotemark[2]\\[2ex]%
	\today
}
			\maketitle
\vspace*{-0.2cm}


\begin{abstract}

In this paper we propose a flexible class of multivariate nonlinear non-Gaussian state space models, based on copulas.
More precisely, we assume that the observation equation and the state equation are defined by copula families that are not necessarily equal.
For each time point, the resulting model can be described by a C-vine copula truncated after the first tree, where the root node is represented by the latent state.
Inference is performed within the Bayesian framework, using the Hamiltonian Monte Carlo method, where a further D-vine truncated after the first tree is used as prior distribution to capture the temporal dependence in the latent states.
Simulation studies show that the proposed copula-based approach is extremely flexible, since it is able to describe a wide range of dependence structures and, at the same time, allows us to deal with missing data.
The application to atmospheric pollutant measurement data shows that our approach is suitable for accurate modeling and prediction of data dynamics in the presence of missing values.
Comparison to a Gaussian linear state space model and to Bayesian additive regression trees shows the superior performance of the proposed model with respect to predictive accuracy.

\end{abstract}

\noindent%

\section{Introduction}\label{intro}

State space models, also called dynamic models, originated in the field of system theory and were introduced by \cite{kalman1960new} and \cite{kalman1961new}, with early applications in aerospace-related research (\cite{hutchinson1984kalman}).
Since then, state space models have gained popularity in a number fields and have been applied in different areas, such as economics (\cite{kitagawa1984smoothness, shumway1982approach}), medicine (\cite{myers2007state, liu2015modeling}) and ecology (\cite{fruhwirth1994data}).
\cite{durbin2000time, durbin2002simple, durbin2012time} provide a thorough illustration of state space models for time series analysis.

Linear Gaussian state space models are the most popular models in this class, with several contributions in the literature, including, for example, \cite{ippoliti2012space} who applied these approaches to environmental data and \cite{van2010intervention}, to official statistics.
However, the strong assumptions of linear Gaussian state space models prevent their applicability to data showing departures from linearity and normality. 
In order to overcome these limitations, \cite{chen2012tracking} applied nonlinear state space models to an epidemiological study on measles infection, relaxing the linearity assumptions, yet assuming normality for the model equations.
\cite{johns2005non} proposed a spatio-temporal model for the analysis of censored dust particle concentrations which overcomes the linearity and normality assumptions, but assumes conditional Gaussian equation errors. 

Copula-based approaches have proven to be particularly suitable for modeling data showing departures from multivariate normality.
Copulas allow us to model separately the marginals from the dependence structure, and the use of different copula families, particularly Archimedean copulas such as the Clayton and Gumbel, are suitable to accommodate asymmetric tail dependence.
The literature of copula applications is vast. For environmental, actuarial and financial applications, see, for example, \cite{genest2007everything}, \cite{patton2006modelling}, \cite{jondeau2006copula}, \cite{cherubini2004copula}, among others. A detailed overview of copulas and their properties is given by \cite{joe1997multivariate} and \cite{nelsen2007introduction}.  

A rich class of parametric copula families is available in the bivariate case.
However, in higher dimension the applicability of copulas is mostly limited in practice to the multivariate Gaussian or Student t.
Vines, constructed using bivariate copulas as building blocks, provide a flexible alternative in the multivariate case.
Vine copulas were first introduced by \cite{joe1996families} and organised in a systematic way using graphical model structures by \cite{bedford2002vines}. A thorough introduction to vines is provided by \cite{aas2009pair} and \cite{czado2019analyzing}.
Special types of vine copulas are C-vines, where one variable plays the role of the root node in each level, and D-vines, constructed as sequences of bivariate copulas.
D-vines were employed by \cite{smith2010modeling} to model the dependence structure of longitudinal data.

\cite{hafner2012dynamic} and \cite{almeida2012efficient} suggest a bivariate state space model, with a bivariate copula in the observation equation and a Gaussian autoregressive process of order one, which describes the time evolution of the copula parameter, in the state equation.
 \cite{kreuzer2019bayesian} propose a univariate nonlinear non-Gaussian state space model, where both the observation and the state equation, are defined in terms of copula specifications. 
 However, the copulas describing the observation and the state equation belong to the same family.

We propose a multivariate nonlinear non-Gaussian state space model, which extends the approach introduced by \cite{kreuzer2019bayesian} to multivariate observations, which we assume to be related to an underlying latent variable.
This approach allows us to capture cross-sectional as well as temporal dependence in a very flexible way, since the copulas specifying the model can be all different.
For each time point, the proposed model can be described as a C-vine truncated at the first tree, with the latent state being the root node. 
The latent states are treated as parameters, with prior distribution given by a D-vine truncated after the first tree to capture temporal dependence. An advantage of our approach is that missing values are handled in a natural way, since they are treated as latent variables.
For model estimation, we cannot rely on the standard Kalman filter approach developed for linear state space models.
Therefore, we suggest a Bayesian approach implemented using the Hamiltonian Monte Carlo (HMC) method (\cite{neal2011mcmc}, \cite{carpenter2017stan}), where we introduce an indicator variable for the copula families specifying the state space model equations. 

We demonstrate the usefulness of our method in a data set containing different air pollutant measurements. Three different pollutants are considered, and for each pollutant, measurements from a high-cost and from a low-cost sensor are utilized. In addition, covariates such as the temperature are available. To model this data we follow a flexible two-step modeling approach, motivated by Sklar's Theorem (\cite{sklar1959fonctions}). First we model the marginal distributions with generalized additive models (\cite{hastie1987generalized}) and in the second step we model dependencies with the novel copula state space model. 
We utilize our model to reconstruct high-cost measurements from low-cost measurements as in \cite{de2008field} and show that the copula-based state space model, in combination with marginal generalized additive models, does a good job at predicting high-cost measurements. We show that it outperforms a Gaussian state space model and Bayesian additive regression trees with respect to the continuous ranked probability score (\cite{gneiting2007strictly}).

The rest of the paper is organized as follows: Section \ref{CopulaStateSpace} introduces the novel multivariate copula state space model, 
Section \ref{sec:bayesinf} discusses Bayesian inference for the novel approach, Section \ref{sec:dataana} is devoted to the air pollutant measurements application and Section \ref{Conclusions} concludes.

\section{The Model}\label{CopulaStateSpace}

Copula approaches are very flexible since they can be combined with different marginal distributions. 
For the air pollution measurements data with additional covariates, as analyzed in Section \ref{sec:dataana}, we propose 
generalized additive models (GAMs) for the margins in combination with the novel copula state space model to capture dependencies. 
The GAM explains the effect of the covariates, while the copula-based state space model handles temporal and cross-sectional dependence. In this section, we first introduce the marginal models (Section \ref{marginal}), which 
yield data on the copula scale. Then, we review the linear Gaussian state space model (Section \ref{sec:lingauss}) and show an equivalent formulation in terms of Gaussian copulas (Section \ref{sec:gausscopform}). In Section \ref{sec:copstate_def} we finally introduce the multivariate copula state space model as a generalization of the linear Gaussian state space model. The behavior of this model is illustrated with simulated data in Section \ref{sec:illus}.

\subsection{Marginal Models}\label{marginal}

Consider a multivariate time series $\textbf{Y}_{t} = (Y_{t1}, \ldots, Y_{td})'$ corresponding to $d$-dimensional continuous data, observed at the time points $t = 1, \ldots, T$, that may depend on a $q$-dimensional covariate vector $\boldsymbol{x}_t = (x_{t1}, \ldots, x_{tq})'$. 

In order to allow for more flexibility, we consider Box-Cox transformations  (\cite{box1964analysis}) of the response variables, i.e. we  consider the transformed variables

\begin{equation}
BC(Y_{tj},\lambda_j) =\begin{cases}
\frac{Y_{tj}^{\lambda_{j}} - 1}{\lambda_j}, \mbox{  for } \lambda_j \neq 0\\
\ln(Y_{tj}),  \mbox{  for }  \lambda_j = 0\\
\end{cases}.
\label{eq:boxcox}
\end{equation}

The relationship between the Box-Cox-transformed variables and the covariates can be expressed in various ways, e.g. using linear or nonlinear regression models.
We assume a GAM (\cite{hastie1987generalized})
 such that
$$
BC(Y_{tj}, \lambda_j) = f_j (\boldsymbol{x}_t) + \sigma_j \varepsilon_{tj},
$$
where $f_j(\cdot)$ is a smooth function of the covariates, expressing the mean of the GAM, and $\varepsilon_{tj} {\sim} N(0,1)$.
Let us define the standardized errors of the GAM as
\begin{equation}
Z_{tj} = \frac{BC(Y_{tj}, \lambda_j) - f_j (\boldsymbol{x}_t)}{\sigma_j}.
\label{eq:gamerr}
\end{equation}
Note that $Z_{tj} {\sim} N(0,1)$ holds.

We aim at modeling the errors $\textbf{Z}_{t} = (Z_{t1}, \ldots, Z_{td})'$ as a multivariate nonlinear non-Gaussian state space model based on copulas.

\subsection{Linear Gaussian State Space Models}
\label{sec:lingauss}

State space models relate observations of a response variable to unobserved latent variables or ``states''. Gaussian linear state space models are defined by a linear observation model and
a linear Markovian transition equation (\cite{durbin2000time}, \cite{durbin2002simple}, \cite{durbin2012time}).

Suppose that we model the errors $\textbf{Z}_{t}$, with $t = 1, \ldots, T$, extracted from the GAM as explained in Section \ref{marginal}, as a linear Gaussian state space model. Here, the variables $Z_{tj}$,  $j = 1, \ldots, d$, are connected to a common continuous state variable $W_t$. Hence, the model can be formulated as
\begin{align}
Z_{tj} &= \rho_{obs, tj} \, W_t + \sigma_{obs, tj} \, \eta_{obs, tj} \label{obs} \\
W_{t} &= \rho_{lat, t} \, W_{t-1} + \sigma_{lat, t} \, \eta_{lat, t}, \label{state}
\end{align}
where $\eta_{obs, tj} \sim N(0,1)$, $\eta_{lat, t} \sim N(0,1)$ are independent i.i.d. sequences, $\rho_{obs, tj}$, $\rho_{lat, t}$, $\sigma_{obs, tj}$ and $\sigma_{lat, t}$ are model parameters and $W_0 \sim N(\mu_{lat, 0}, \sigma_{lat, 0})$, with $\mu_{lat, 0}$ and $\sigma_{lat, 0}$ generally known.
Equation (\ref{obs}) is called observation equation, while Equation (\ref{state}) is called state equation.

The linear Gaussian state space model can also be expressed using conditional distributions as 
\begin{align}
Z_{tj} \, | \, W_t = w_t & \sim N \left(\rho_{obs, tj} \, w_t; \, \sigma_{obs, tj}^2 \right) \nonumber \\
W_{t} \, | \, W_{t-1} = w_{t-1} & \sim N \left(\rho_{lat, t} \, w_{t-1}; \,  \sigma_{lat, t}^2 \right). \nonumber
\end{align}

We assume time stationarity, i.e. $\rho_{obs, tj} = \rho_{obs, j}$, for $j = 1, \ldots, d$, and $\rho_{lat, t} = \rho_{lat}$. Since the model is applied to standardized errors with unit variance we also set $\sigma_{obs, tj}^2=1-\rho_{obs, j}^2$ and  $\sigma_{lat, t}^2=1-\rho_{lat}^2$.   In addition, we assume that $\mu_{lat, 0} = 0$ and $\sigma_{lat, 0} = 1$. These assumptions imply that $Z_{tj} \sim N(0,1)$ unconditionally.
Hence, the model expressed through conditional distributions becomes
\begin{align}
Z_{tj} \, | \, W_t = w_t & \sim N \left(\rho_{obs, j} \, w_t; \, 1-\rho_{obs, j}^2 \,  \right) \nonumber \\
W_{t} \, | \, W_{t-1} = w_{t-1} & \sim N \left(\rho_{lat} \, w_{t-1}; \,  1-\rho_{lat}^2 \, \right). \nonumber
\end{align}
Thus, the state space model induces the following bivariate Gaussian distribution
\begin{equation}\label{obsGauss}
\begin{pmatrix} 
   Z_{tj} \\
    W_t \\
\end{pmatrix} \sim N_2\left( \begin{pmatrix}
   0 \\
    0 \\
\end{pmatrix} ,   \begin{pmatrix}
    1       & \rho_{obs, j} \\
    \rho_{obs, j}       & 1 \\
\end{pmatrix}\right) \\
\end{equation}
\begin{equation}\label{latGauss}
\begin{pmatrix}
   W_{t} \\
    W_{t-1} \\
\end{pmatrix} \sim N_2\left( \begin{pmatrix}
   0 \\
    0 \\
\end{pmatrix} ,   \begin{pmatrix}
    1       & \rho_{lat} \\
    \rho_{lat}       & 1 \\
\end{pmatrix}\right). \\
\end{equation}
Therefore, we obtain the joint distribution 
$$
(Z_{11}, \ldots, Z_{d1}, W_1; Z_{12}, \ldots, Z_{d2}, W_2; \ldots, Z_{1T}, \ldots, Z_{dT}, W_T) \sim N_{(d+1)T} (\textbf{0}, \Sigma)
$$
with covariance matrix $\Sigma$ (see supplementary material).
Thus, the joint distribution of $Z_{tj}$ and $Z_{t-1 j}$ is given by
$$
\begin{pmatrix}
   Z_{tj} \\
    Z_{t-1 j} \\
\end{pmatrix} \sim N_2\left( \begin{pmatrix}
   0 \\
    0 \\
\end{pmatrix} ,   \begin{pmatrix}
    1       & \rho_{obs,j}^2 \rho_{lat} \\
    \rho_{obs,j}^2 \rho_{lat}       & 1 \\
\end{pmatrix}\right). \\
$$

\subsection{Copula Formulation of a Gaussian State Space Model}
\label{sec:gausscopform}

The linear Gaussian state space model in equations (\ref{obsGauss}) and (\ref{latGauss}) can be equivalently expressed in the copula space using Gaussian copulas as follows
\begin{equation}
\begin{split}
(U_{tj} \, , \, V_t)  & \sim  \mathbbm{C}_{U_j,V}^{Gauss} (\, \cdot \,, \, \cdot; \, \tau_{obs, j})  \\
(V_t \, , \, V_{t-1})  & \sim  \mathbbm{C}_{V_{2},V_{1}}^{Gauss} (\, \cdot \, , \, \cdot; \, \tau_{lat}), 
\label{eq:gaus_cop_ssm}
\end{split}
\end{equation}
where 
\begin{equation} \label{eq:residuals}
    U_{tj} = \Phi \left( Z_{tj} \right), \, \, \, V_t = \Phi \left( W_t \right),  \, \, \, \, \, j = 1, \ldots, d,  \, \, \, \, \,  t = 1, \ldots, T ,
\end{equation}
with $\Phi$ denoting the standard normal cumulative distribution function.
The variables $U_{tj}$ and $V_t$ are uniformly distributed as  $U_{tj} \sim U \left( 0, 1 \right), \, \, \, V_t \sim U \left( 0, 1 \right)$,
while the variables $Z_{tj}$ and $W_t$ are normally distributed as 
$Z_{tj} \sim N \left( 0, 1 \right), \, \, \, W_t \sim N \left( 0, 1 \right)$.
The Gaussian copulas in (\ref{eq:gaus_cop_ssm}) are parametrized by Kendall's $\tau$, such that  $\tau_{obs,j} =  \frac{2}{\pi} \arcsin(\rho_{obs, j}), \, \, \, \tau_{lat} =  \frac{2}{\pi} \arcsin(\rho_{lat})$.

\subsection{Multivariate Nonlinear Non-Gaussian Copula State Space Model}
\label{sec:copstate_def}

The multivariate nonlinear non-Gaussian copula state space model allows the copula families in (\ref{eq:gaus_cop_ssm}) to be different from the Gaussian, thus gaining a much greater flexibility to accommodate a wide range of dependence structures.   

More precisely, the proposed model can be expressed, in the copula scale, as follows
\begin{equation}
\begin{split}
(U_{tj} \, , \, V_t)  & \sim  \mathbbm{C}_{U_j,V}^{m_{obs,j}} (\, \cdot \,, \, \cdot; \, \tau_{obs, j})  \\
(V_t \, , \, V_{t-1})  & \sim  \mathbbm{C}_{V_{2},V_{1}}^{m_{lat}} (\, \cdot \, , \, \cdot; \, \tau_{lat}), 
\label{eq:cop_cop_ssm}
\end{split}
\end{equation}
where the copula families $m_{obs,j}$, for $j = 1, \ldots, d$, and $m_{lat}$ are not necessarily equal and belong to a set $\mathcal{M}$ of single parameter copula families, parametrized by $\tau_{obs, j} = g_{m_{obs,j}}(\theta_{obs, j}^{m_{obs,j}})$ and $\tau_{lat} = g_{m_{lat}}(\theta_{lat}^{m_{lat}})$. The functions $g_{m_{obs,j}}, g_{m_{lat}}$ are one-to-one transformation functions and $\theta_{obs, j}^{m_{obs,j}}$ and $\theta_{lat}^{m_{lat}}$ are the parameters of the bivariate copulas $\mathbbm{C}_{U_j,V}^{m_{obs,j}}$ and $\mathbbm{C}_{V_{2},V_{1}}^{m_{lat}}$, respectively. For example, for the Gumbel copula 
$g_{Gumbel}(\theta_{obs, j}^{Gumbel}) = 1 -\frac{1}{\theta_{obs, j}^{Gumbel}}$ holds.

The proposed model can also be specified in terms of conditional distribution functions as follows
\begin{equation}
\begin{split}
(U_{tj} \, | \, V_t = v_t)  & \sim  \mathbbm{C}_{U_j|V}^{m_{obs,j}} (\, \cdot \,| \, v_t; \, \tau_{obs, j})  \\
(V_t \, | \, V_{t-1} = v_{t-1})  & \sim  \mathbbm{C}_{V_{2}|V_{1}}^{m_{lat}} (\, \cdot \, | \, v_{t-1}; \, \tau_{lat}), 
\label{eq:gen_cop_ssm}
\end{split}
\end{equation}

\tikzstyle{observed} = [circle, draw, thin, minimum height=2.5em]
\tikzstyle{latent} = [draw, thin, minimum height=2.5em]
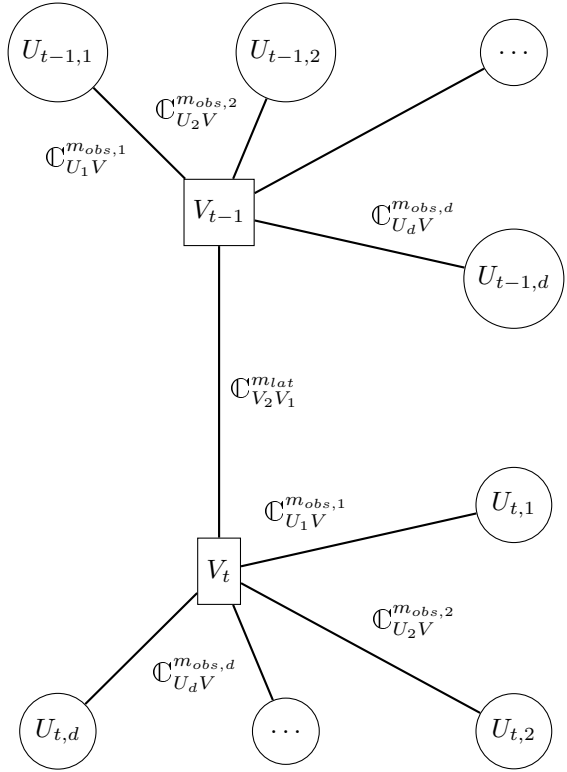
\begin{figure}[H]
\centering
\begin{tikzpicture}[node distance=3cm, auto,>=latex', thick]
    \path[-] node[latent] (vt1) {$V_{t-1}$};
    \path[-] node[observed, above left of=vt1] (u1t1) {$U_{t-1, 1}$}
    				(vt1) edge node {$\mathbbm{C}_{U_{1}V}^{m_{obs,1}}$} (u1t1);  
    \path[-] node[observed, right of=u1t1] (u2t1) {$U_{t-1, 2}$}
    				(vt1) edge node {$\mathbbm{C}_{U_{2}V}^{m_{obs,2}}$} (u2t1);  
    \path[-] node[observed, right of=u2t1] (dots1) {$\ldots$}
    				(vt1) edge node {} (dots1);
     \path[-] node[observed, below of=dots1] (upt1) {$U_{t-1, d}$}
    				(vt1) edge node {$\mathbbm{C}_{U_{d}V}^{m_{obs,d}}$} (upt1);	
    \path[-] node[observed, below of=upt1] (u1t) {$U_{t, 1}$};  
    \path[-] node[observed, below of=u1t] (u2t) {$U_{t, 2}$};	 
    \path[-] node[observed, left of=u2t] (dots2) {$\ldots$};				
    \path[-] node[observed, left of=dots2] (upt) {$U_{t, d}$};
	\path[-] node[latent, above right of=upt] (vt) {$V_t$}
                  (vt1) edge node {$\mathbbm{C}_{V_{2}V_{1}}^{m_{lat}}$} (vt)
					(vt) edge node {$\mathbbm{C}_{U_{1}V}^{m_{obs,1}}$} (u1t)
					(vt) edge node {} (dots2)
					(vt) edge node {$\mathbbm{C}_{U_{2}V}^{m_{obs,2}}$} (u2t)
					(vt) edge node {$\mathbbm{C}_{U_{d}V}^{m_{obs,d}}$} (upt);   
\end{tikzpicture}
\caption{Graphical visualization of the multivariate state space copula model as specified in \eqref{eq:cop_cop_ssm}.}\label{fig:cop_st_sp}
\end{figure} 

Figure \ref{fig:cop_st_sp} shows a graphical representation of the multivariate state space copula model. Each observed variable $U_{tj}$ is linked to the latent state variable $V_t$ via a copula $\mathbbm{C}_{U_{j}V}^{m_{obs,j}}$ and the dependence between the latent states is modeled by the copula $\mathbbm{C}_{V_{2}V_{1}}^{m_{lat}}$.
In the following we denote by $c_{U_j,V}^{m_{obs,j}}$ and $c_{V_{2},V_{1}}^{m_{lat}}$ the density functions of $\mathbbm{C}_{U_j,V}^{m_{obs,j}}$ and $\mathbbm{C}_{V_{2},V_{1}}^{m_{lat}}$, respectively.

\subsection{Illustration of the Copula State Space Model with Simulated Data}
\label{sec:illus}

\begin{figure}[H]
\centerline{%
\includegraphics[trim={0 1cm 0 0},width=0.95\textwidth]{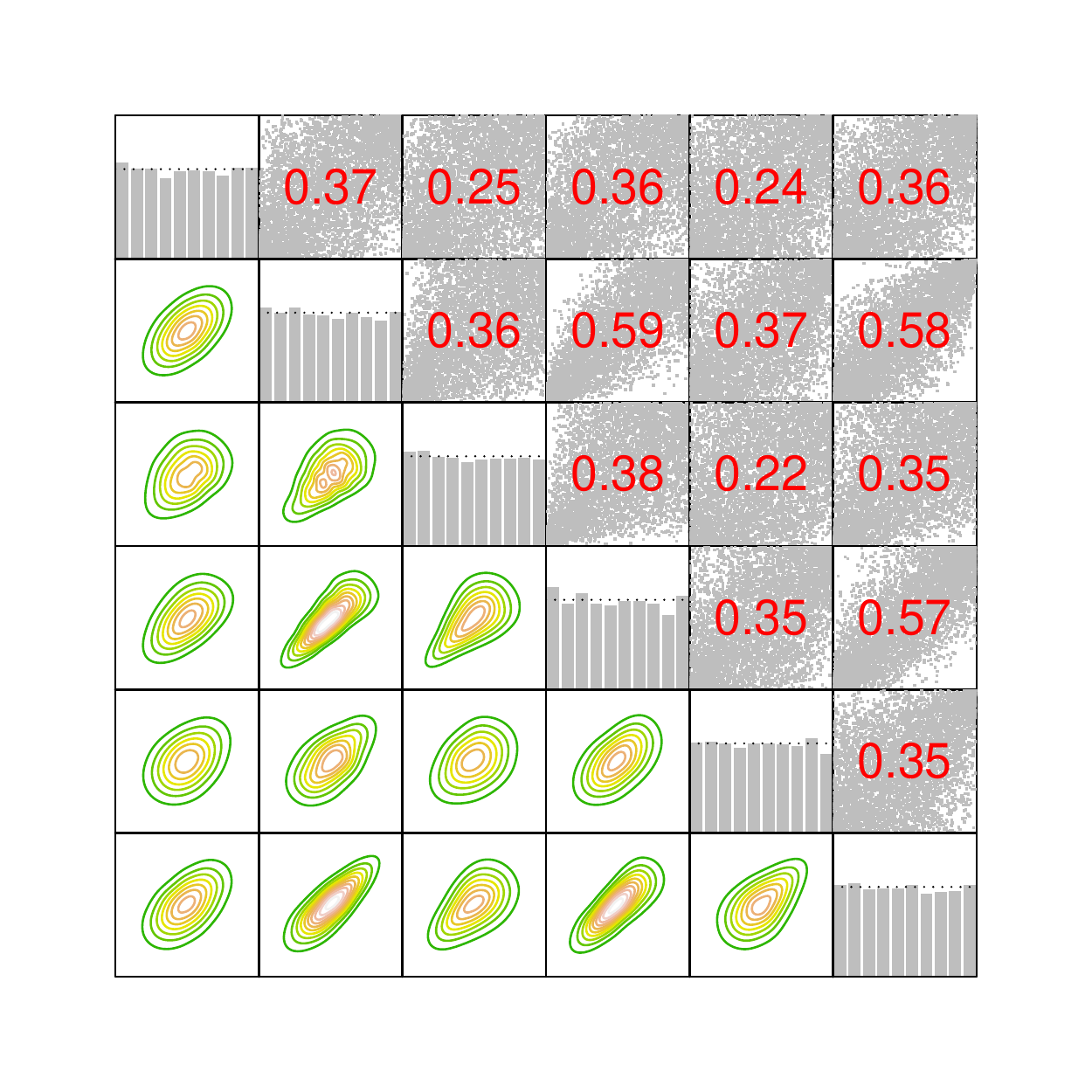}%
}%
\caption{This plot is based on independently simulated data $(u_{tj}^r)_{r=1, \ldots, 5000, j=1, \ldots, 6}$ from Scenario 1 for a fixed $t$ $\in$ $\{1, \ldots, T\}$. The lower triangular part shows contour plots of all pairs of $(z_{t1}^r, \ldots,  z_{t6}^r), r=1, \ldots, 5000$, where $z_{tj}^r = \Phi^{-1}(u_{tj}^r)$. The upper triangular part shows corresponding scatter plots and the empirical Kendall's $\tau$ for each pair $(u_{tj},u_{tj'})$. The diagonal shows the histogram of the univariate marginals. More precisely, the plot in the $i$-th row and $j$-th column shows the contour plot for the pair $( z_{ti}^r,  z_{tj}^r)$ if $i>j$, the scatter plot of $( u_{ti}^r,  u_{tj}^r)$ if $i<j$, or the histogram of $ u_{ti}^r$, if $i=j$, with $r=1, \ldots, 5000$.}
\label{fig:simcont1}
\end{figure}

We visualize bivariate dependence structures that are obtained from our model with normalized contour plots (see, for example, \cite{czado2019analyzing}, Chapter 3). We consider three scenarios which differ in the choice of the family $m_{lat}$ of the latent copula. The parameters are chosen as follows
\begin{equation}
\begin{split}
& T=1000 \\
& d=6 \\
&\boldsymbol  {m_{obs}}  = (\text{Gaussian, Gaussian, Clayton, Clayton, Gumbel, Gumbel}) \\
&\boldsymbol  {\tau_{obs}} = (0.5,0.7,0.5,0.7,0.5,0.7) \\
& \tau_{lat} = 0.7 \\
& m_{lat} = \begin{cases} \text{Gaussian, Scenario 1} \\
\text{Clayton, Scenario 2 }\\
\text{Gumbel, Scenario 3} \end{cases}
\end{split}
\end{equation}
We consider one symmetric bivariate copula (Gaussian) and two asymmetric bivariate copulas (Gumbel, Clayton).
We investigate two types of dependence: cross-sectional and temporal. For the cross-sectional dependence, we consider the pairs $(U_{tj}, U_{tj'})$ with corresponding bivariate copula density
\begin{equation}
c(u_{tj},u_{tj'}) = \int_0^1 c^{m_{obs,j}}_{U_jV}(u_{tj},v_t)  c^{m_{obs,j'}}_{U_{j'}V}(u_{tj'},v_t) dv_t.
\label{eq:marg_ucross}
\end{equation}
The bivariate marginal density of $(U_{tj}, U_{tj'})$ given in \eqref{eq:marg_ucross} is neither affected by the time $t$ nor by the copula $\mathbbm{C}^{m_{lat}}_{V_2V_1}$. So the cross-sectional dependence is not affected by the copula $\mathbbm{C}^{m_{lat}}_{V_2V_1}$ and the corresponding theoretical contour plots are the same for all three scenarios. The empirical normalized contour plots for pairs $(U_{tj}, U_{tj'})$ are shown in Figure \ref{fig:simcont1} for Scenario 1. The contour plots are constructed from 5000 independent simulations of the density in \eqref{eq:marg_ucross} for a fixed $t \in \{1, \ldots, T\}$.

 We see that if both linking copulas $\mathbbm{C}^{m_{obs,j}}_{U_jV}$ and $\mathbbm{C}^{m_{obs,j'}}_{U_{j'}V}$ are Gaussian, the contour of $(U_{tj}, U_{tj'})$ looks Gaussian as well (
 see the panel in the second row and the first column in Figure \ref{fig:simcont1}). In this case $\mathbbm{C}(u_{tj},u_{tj'})$ is indeed a Gaussian copula. If we mix a Gaussian and an asymmetric linking copula (
 see the entries below row 2 in columns 1 and 2 in Figure \ref{fig:simcont1}) or if we combine two asymmetric linking copulas (
 see the lower triangular entries in columns 3, 4 and 5 in Figure \ref{fig:simcont1}) we can obtain a variety of different asymmetric contour shapes.

\begin{figure}[H]
\centerline{%
\includegraphics[trim={0 6.25cm 0 0},width=0.8\textwidth]{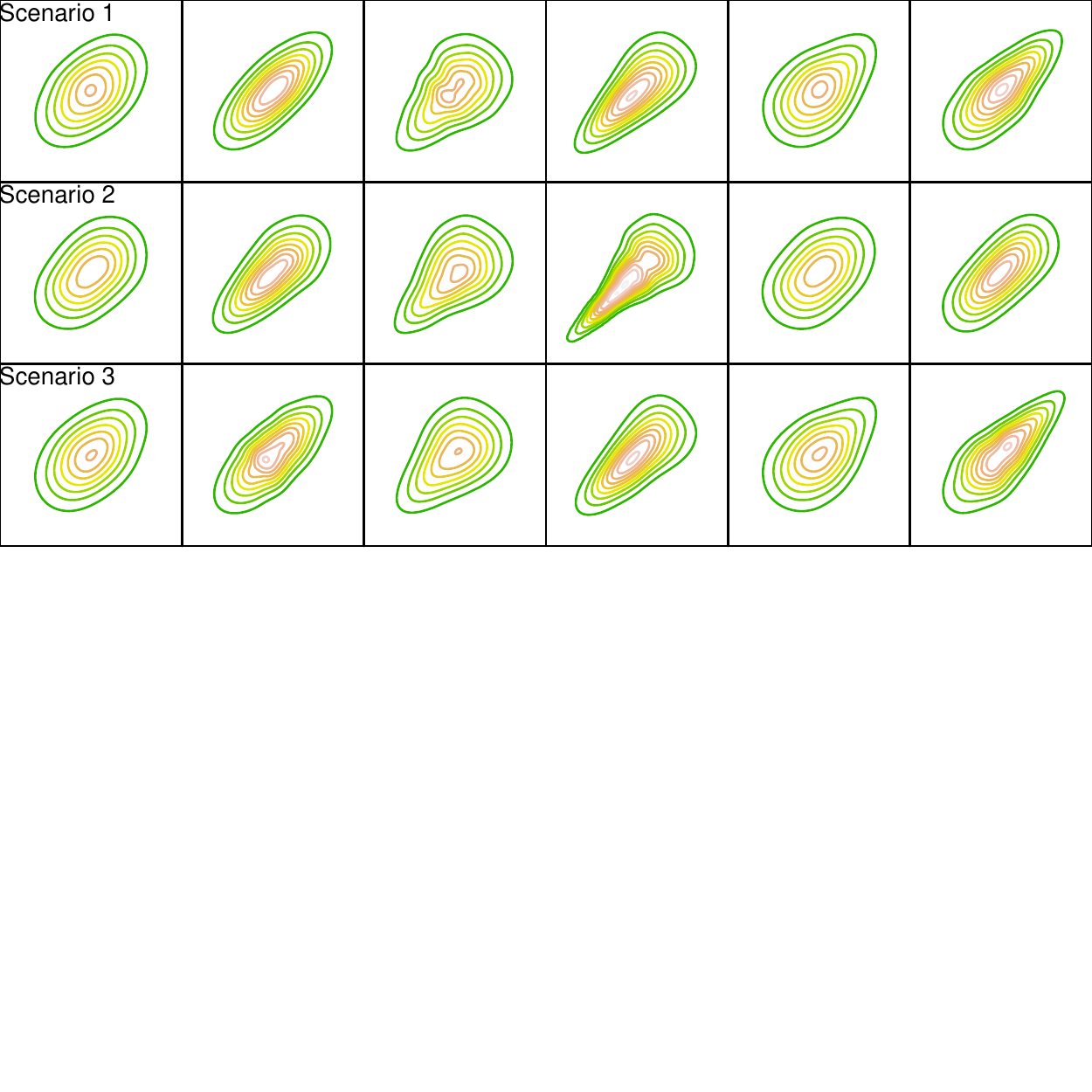}%
}%
\caption{This plot is based on independently simulated data $(u_{t-1j}^r, u_{tj}^r)_{r=1, \ldots, 5000, j=1, \ldots, 6}$ from Scenarios 1--3 for a fixed $t$ $\in$ $\{2, \ldots, T\}$. 
The data is transformed to the normalized scale as $z_{t'j}^r = \Phi^{-1}(u_{t'j}^r)$, $t' = t-1, t$. Contour plots of the pairs $({z}_{tj}^r, {z}_{t-1j}^r)_{r=1, \ldots, 5000}$ are shown for $j=1, \ldots, 6$. The plot in row $m$ and column $j$ shows the contour plot for $({z}_{tj}^r, {z}_{t-1j}^r)_{r=1, \ldots, 5000}$, simulated from the parameter specification of 
Scenario $m$.}
\label{fig:simconttimeall}
\end{figure}

For the temporal dependence, we consider the pairs $(U_{tj},U_{t-1j})$ with bivariate copula density
\begin{equation}
c(u_{tj},u_{t-1j}) = \int_{(0,1)^2} c_{U_jV}^{m_{obs,j}}(u_{tj},v_t) c_{V_2V_1}^{m_{lat}}(v_t,v_{t-1}) c_{U_jV}^{m_{obs,j}}(u_{t-1j},v_t) dv_tdv_{t-1}.
\label{eq:marg_temp}
\end{equation}
This dependence is affected by three copulas. Figure \ref{fig:simconttimeall} shows normalized contour plots of the density in \eqref{eq:marg_temp} obtained from 5000 independent simulations. We can see that if at least one of these copulas is asymmetric we may obtain an asymmetric dependence structure.

\section{Bayesian Inference for the Multivariate Copula State Space Model}
\label{sec:bayesinf}

For the type of data we are dealing with, missing values are common. We denote the set of time indices of observed/non-missing values for dimension $j$ by ${\mathcal{T}_j^{obs}}$ and the set of missing values by $\mathcal{T}_j^{miss} = \{1, \ldots, T\} \setminus \mathcal{T}_j^{obs}, j=1, \ldots, d$. Further, we call $U^{obs} = (u_{tj})_{t \in \mathcal{T}_j^{obs}, j=1, \ldots, d}$ the observed and $U^{miss} = (u_{tj})_{t \in \mathcal{T}_j^{miss}, j=1, \ldots, d}$ the missing values. The missing values can be treated as latent variables. Integrating out the missing values yields the following likelihood for the observed values $U^{obs}$
\begin{equation}
\begin{split}
\ell(\boldsymbol v, \boldsymbol \tau_{obs}, & \boldsymbol m_{obs}|U^{obs}) =
\int_{(0,1)^{|U_{miss}|}} \prod_{j=1}^d \prod_{t=1}^T c^{m_{obs,j}}_{U_jV} (u_{tj}, v_t; \tau_{obs,j}) dU^{miss} = \\ 
&=  \prod_{j=1}^d \left(\prod_{t \in \mathcal{T}_j^{obs}} c^{m_{obs,j}}_{U_jV} (u_{tj}, v_t; \tau_{obs,j})  \prod_{t \in \mathcal{T}_j^{miss}}\int_{(0,1)} c^{m_{obs,j}}_{U_jV} (u_{tj}, v_t; \tau_{obs,j}) du_{tj} \right)  \\ 
&= \prod_{j=1}^d \prod_{t \in \mathcal{T}_j^{obs}} c^{m_{obs,j}}_{U_jV} (u_{tj}, v_t; \tau_{obs,j}).
\end{split}
\label{eq:ll}
\end{equation}
Here the latent variable $V_t$ is treated as a parameter $v_t$, and $\boldsymbol v=(v_1, \ldots, v_T)$, $\boldsymbol {\tau_{obs}}$ $ =$ \newline $(\tau_{obs,1}, \ldots, \tau_{obs,d})$, $\boldsymbol {m_{obs}} = (m_{obs,1}, \ldots, m_{obs,d})$.
In contrast to a complete case analysis, 
information from all observed components is utilized in \eqref{eq:ll}. The last equality in \eqref{eq:ll} uses the fact that in a copula the margins are uniform.

As mentioned above we use a D-vine truncated after the first tree to capture temporal dependence among the latent states, i.e.
\begin{equation}
\pi(\boldsymbol v|\tau_{lat}, m_{lat})  = \prod_{t=2}^T c^{m_{lat}}_{V_2V_1}(v_t, v_{t-1}; \tau_{lat})
\end{equation}
with Kendall's $\tau$ parameter $\tau_{lat}$ and copula family indicator $m_{lat} \in \mathcal{M}$. This is a general Markov model of order 1 and collapses to a Gaussian AR(1) process if the Gaussian copula is used.

We restrict $\tau_{obs,1} \in (0,1)$ to be positive to ensure identifiability. This restriction corresponds to restricting the diagonal entries of the factor loading matrix in conventional Gaussian factor models to be positive (see e.g. \cite{lopes2004bayesian}). For the Kendall's $\tau$ values of the remaining components we use a vague uniform prior on $(-1,1)$, reflecting the fact that we do not have prior knowledge about these quantities. The following prior densities are used
\begin{equation}
\tau_{obs,1} \sim Beta(10,1.5), \,\,\, \tau_{obs,j} \sim U(-1,1), \,\, j=2, \ldots, d, \,\,\, \tau_{lat} \sim U(-1,1).
\label{eq:priorspec}
\end{equation}
For the copula family indicators we use discrete uniform priors, i.e.
\begin{equation}
\pi(m_{obs,j}) = \pi(m_{lat})  = \frac{1}{|\mathcal{M}|}
\end{equation}
for $j=1, \ldots, d$. Further we assume that the Kendall's $\tau$ values and the copula family indicators are a priori independent such that the joint prior density is proportional to
$$
\pi(\boldsymbol {\tau_{obs}}, \boldsymbol {m_{obs}}, \tau_{lat}, m_{lat}, \boldsymbol v) \propto \left(\prod_{t=2}^T c_{V_2V_1}^{m_{lat}}(v_t, v_{t-1}; \tau_{lat})\right) \pi(\tau_{obs,1}),
$$
where $\pi(\tau_{obs,1})$ is the prior density specified in \eqref{eq:priorspec}.
This prior density is a joint density of continuous and discrete parameters. For  discrete parameters $\boldsymbol \delta^{disc}$ and continuous parameters $\boldsymbol \delta^{cont}$ the joint density is defined as
$$
f(\boldsymbol \delta^{cont}, \boldsymbol \delta^{disc}) = f(\boldsymbol \delta^{cont}|\boldsymbol \delta^{disc}) f(\boldsymbol \delta^{disc})
$$
where $f(\boldsymbol \delta^{cont}|\boldsymbol \delta^{disc})$ is a conditional probability density function and $f(\boldsymbol \delta^{disc})$ is a joint probability mass function.

The set of parameters can be summarized as $\mathcal{P}$ =$ \{\tau_{lat}, \boldsymbol {\tau_{obs}},m_{lat}, \boldsymbol {m_{obs}},  \boldsymbol v\}$.
The posterior density of our model is proportional to
\begin{equation}
\begin{split}
f(\mathcal{P}|U^{obs})  \propto &  \left(\prod_{j=1}^d \prod_{t \in \mathcal{T}_j^{obs}} c_{U_jV}^{m_{obs,j}}(u_{tj},v_t, \tau_{obs,j})  \right) \left(  \prod_{t=2}^T c^{m_{lat}}_{V_2V_1}(v_t, v_{t-1};\tau_{lat}) \right)  \pi(\tau_{obs,1}).\\
\end{split}
\label{eq:postdens}
\end{equation}
As in \cite{kreuzer2019bayesian}, sampling from the posterior in \eqref{eq:postdens} is not straightforward, e.g. Kalman filter recursions cannot be applied. Since the No-U-turn sampler of \cite{hoffman2014no} has shown good performance for the univariate copula state space model (\cite{kreuzer2019bayesian}), we also use it here. The No-U-Turn sampler is an extension of Hamiltonian Monte Carlo (HMC, \cite{neal2011mcmc}) with adaptively selected tuning parameters. To run the sampler we use STAN (\cite{carpenter2017stan}).

\subsubsection*{Updating Continuous Parameters}
Since HMC cannot deal with discrete variables we integrate over the discrete family indicators which corresponds to summing over them, i.e. 
\begin{equation}
\begin{split}
f(\tau_{lat}, \boldsymbol {\tau_{obs}}, \boldsymbol v|U^{obs}) =& \sum_{ (m_{lat},\boldsymbol{m_{obs}}) \in \mathcal{M}^{d+1}}     f(\tau_{lat}, \boldsymbol {\tau_{obs}},m_{lat}, \boldsymbol {m_{obs}},  \boldsymbol v|U^{obs})\\
\propto&\prod_{j=1}^d \left( \sum_{m_{obs,j} \in \mathcal{M}} \prod_{t \in \mathcal{T}_j^{obs}} c_{U_jV}^{m_{obs,j}}(u_{tj},v_t; \tau_{obs,j})  \right) \cdot \\ & \cdot  \left(\sum_{m_{lat} \in \mathcal{M}}\prod_{t=2}^T c^{m_{lat}}_{V_2V_1}(v_t, v_{t-1};\tau_{lat})   \right) \pi( \tau_{obs,1})
\end{split}
\label{eq:contpar}
\end{equation}
To sample from this density we use STAN's No-U-Turn sampler. 

\subsubsection*{Updating the (Discrete) Copula Family Indicators}
In $f(\boldsymbol {m_{obs}}, m_{lat}|\tau_{lat}, \boldsymbol {\tau_{obs}}, \boldsymbol v, U^{obs})$, all components of $(\boldsymbol {m_{obs,}}, m_{lat})$ are independent. We have that
\begin{equation}
\begin{split}
f( m_{obs,j}|\tau_{lat}, & \boldsymbol \tau_{obs}, \boldsymbol v,\boldsymbol {m_{obs,-j}}, m_{lat}, U^{obs})  =\\ 
&\frac{f(\tau_{lat}, \boldsymbol {\tau_{obs}},m_{lat}, \boldsymbol {m_{obs}},  \boldsymbol v|U^{obs})}{\sum_{m_{obs,j}' \in \mathcal{M}} f(\tau_{lat}, \boldsymbol {\tau_{obs}},m_{lat}, \boldsymbol {m_{obs,-j}},m_{obs,j}',  \boldsymbol v|U^{obs})}, \\
\end{split}
\end{equation}
where $\boldsymbol {m_{obs,-j}}$ is equal to $\boldsymbol {m_{obs}}$ with the $j$-th component removed. Therefore we obtain
\begin{equation}
\begin{split}
f( m_{obs,j}|\tau_{lat}, \boldsymbol {\tau_{obs}}, \boldsymbol v,\boldsymbol {m_{obs,-j}}, m_{lat}, U^{obs}) & = \frac{\prod_{t \in \mathcal{T}_j^{obs}} c_{U_jV}^{m_{obs,j}}(u_{tj},v_t; \tau_{obs,j})}{\sum_{m_{obs,j}' \in \mathcal{M}}\prod_{t \in \mathcal{T}_j^{obs}} c_{U_jV}^{m_{obs,j}'}(u_{tj},v_t; \tau_{obs,j})} \\
\end{split}
\label{eq:cdist_mobs}
\end{equation}
Similarly we obtain
\begin{equation}
\begin{split}
f( m_{lat}|\tau_{lat}, \boldsymbol {\tau_{obs}}, \boldsymbol v,\boldsymbol {m_{obs}}, U^{obs}) & = \frac{\prod_{t=2}^T c_{V_2V_1}^{m_{lat}}(v_{t},v_{t-1}; \tau_{lat})}{\sum_{m_{lat}' \in \mathcal{M}}\prod_{t=2}^T c_{V_2V_1}^{m_{lat}'}(v_t,v_{t-1}; \tau_{lat})} \\
\end{split}
\label{eq:cdist_mlat}
\end{equation}

\subsubsection*{Obtaining Updates for the Joint Posterior Density}
To obtain $R$ samples from the posterior density given in \eqref{eq:postdens} we first obtain $R$ samples of $\tau_{lat}, \boldsymbol {\tau_{obs}}, \boldsymbol v$ from the density given in \eqref{eq:contpar} using STAN. We denote the samples by $\tau_{lat}^r$, $\boldsymbol {\tau_{obs}^r}$, $\boldsymbol {v^r}$, $r=1, \ldots, R$. Then we sample $m_{obs,j}$ from $f( m_{obs,j}|\tau_{lat}^r, \boldsymbol {\tau_{obs}^r}, \boldsymbol {v^r}, U^{obs})$ (see \eqref{eq:cdist_mobs}) to obtain $m_{obs,j}^r$, for $r=1, \ldots, R$ and $j=1, \ldots, d$. Further, $m_{lat}^r$ is obtained by sampling from $f( m_{lat}|\tau_{lat}^r, \boldsymbol {\tau_{obs}^r}, \boldsymbol {v^r}, U^{obs})$ (see \eqref{eq:cdist_mlat}) , for $r=1, \ldots, R$.  

\subsubsection*{Predictive Distribution (In-Sample Period)}
\label{sec:pdist_ins}
The predictive density of a new value $u_{tj}^{new}$ for margin $j$ at time $t \in \{1, \ldots, T\}$ is the conditional density of $u_{tj}^{new} $ given $U^{obs}$, obtained as
$$
f(u_{tj}^{new}|U^{obs}) = \int_{domain(\mathcal{P})} f(u_{tj}^{new}, \mathcal{P}| U^{obs}) d\mathcal{P} = \int_{domain(\mathcal{P})} f(u_{tj}^{new}|\mathcal{P}, U^{obs}) f(\mathcal{P}|U^{obs})  d\mathcal{P}
$$
with  $f(u_{tj}^{new}| \mathcal{P}, U^{obs})= c^{m_{obs,j}}_{U_jV}(u_{tj}^{new},v_t;\tau_{obs,j})$ and $domain(\mathcal{P})$ is the domain of the parameter space $\mathcal{P}$.
Note that for the discrete indicator variables the integral is a sum.

To obtain samples from the predictive distribution we sample from the following density 
$$
f(u_{tj}^{new}, \mathcal{P}| U^{obs}) = f(u_{tj}^{new}|\mathcal{P}, U^{obs}) f(\mathcal{P}|U^{obs}).
$$
We proceed as follows:
\begin{itemize}
\item We first simulate $R$ samples of $\mathcal{P}$ from $f(\mathcal{P}|U^{obs})$ as described above.
\item The $r-th$ sample of $u^{new}_{tj}$, denoted by $(u^{new}_{tj})^r$, is simulated from $\mathbbm{C}_{U_j|V}^{m_{obs,j}^r}(\cdot|v_t^r;\tau_{obs,j}^r)$, for $r=1, \ldots, R$.
\end{itemize}
For $t \in  \mathcal{T}_j^{miss}$, we can obtain simulated values for the missing values.

\subsubsection*{Predictive Distribution (Out-of-Sample Period)}
\label{sec:pdist_oos}

To obtain samples from the predictive distribution of a new value $u_{tj}^{new}$ for margin $j$ at time $t \in \{T+1, T+2, \ldots\}$  we consider the following density
$$
f(u_{tj}^{new}, \mathcal{P}|U^{obs}) = f(u_{tj}^{new}|\mathcal{P}, U^{obs}) f(\mathcal{P}|U^{obs}) 
$$
with  $$f(u_{tj}^{new}| \mathcal{P}, U^{obs})= \int_{(0,1)^{t-T}} c^{m_{obs,j}}_{U_jV}(u_{tj}^{new},v_t;\tau_{obs,j}) \prod_{t'=T+1}^{t}c^{m_{lat}}_{V_2V_1}(v_{t'},v_{t'-1};\tau_{lat}) dv_{T+1}\ldots, dv_t.$$
We proceed as follows to obtain samples from this density
\begin{itemize}
\item We first simulate $R$ samples of $\mathcal{P}$ from $f(\mathcal{P}|U^{obs})$ as described above. 
\item For $r=1, \ldots, R$ and  for $t'=T+1, \ldots, t$: \newline
Sample $v_{t'}$ from $\mathbbm{C}^{m_{lat}^r}_{V_2V_1}(\cdot|v_{t'-1}^r;\tau_{lat}^r)$ and denote the sample by $v_{t'}^r$.
\item For $r=1, \ldots, R$: Sample $u_{tj}^{new}$ from $\mathbbm{C}_{U_j|V}^{m_{obs,j}^r}(\cdot|v_t^r;\tau_{obs,j}^r)$ and denote the sample by $(u^{new}_{tj})^r$.
\end{itemize}
Note that the recursive sampling avoids the evaluation of the $t-T$ dimensional integral.

\section{Data Analysis}
\label{sec:dataana}
\subsection{Data Description}
\begin{figure}[H]
\centerline{%
\includegraphics[trim={0 11.25cm 0 0},width=1.0\textwidth]{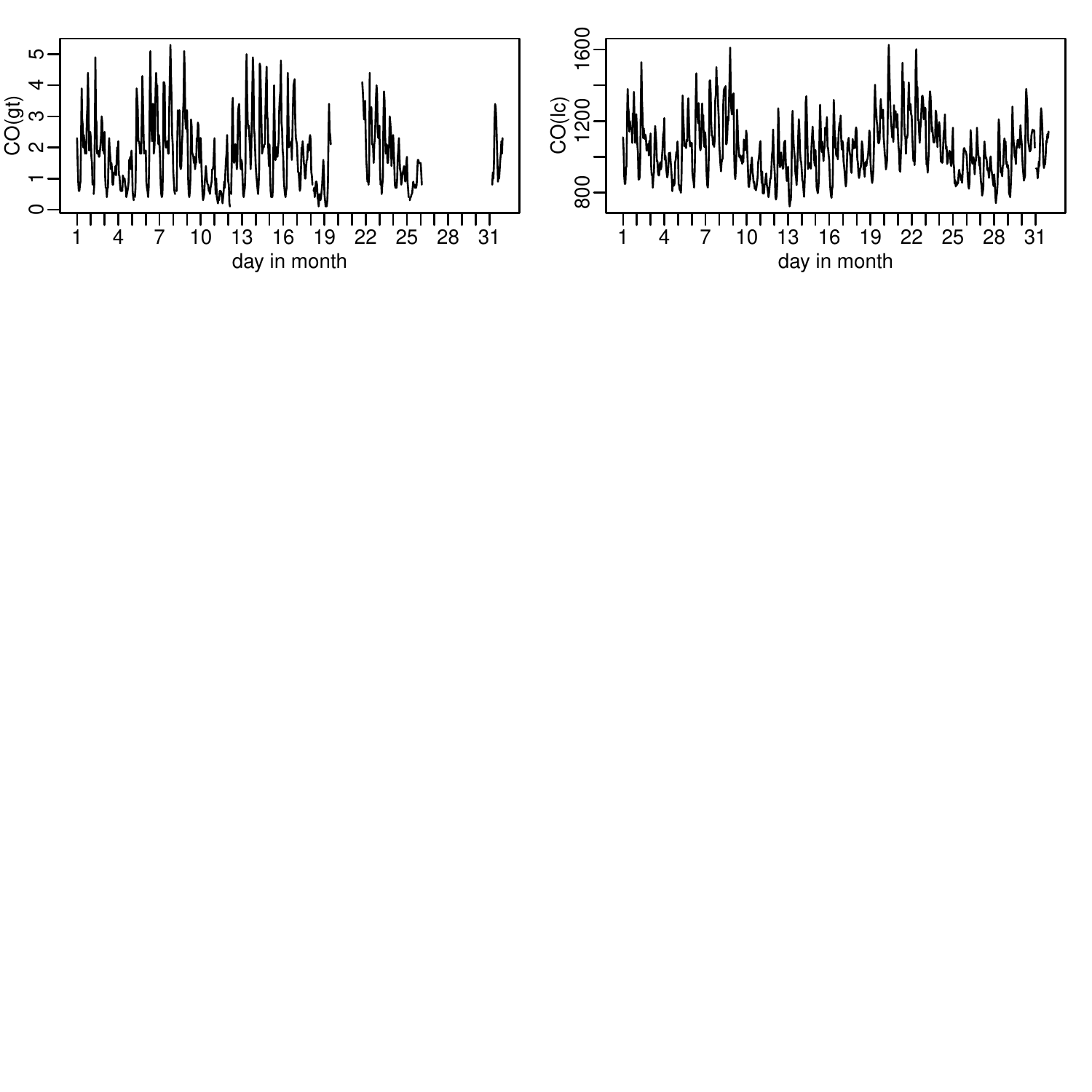}
}%
\caption{Hourly observed values of one pollutant (CO) from the ground truth (gt) and low-cost (lc) sensors in July 2004. When missing values are present no observations are drawn for the corresponding time points.}
\label{fig:data}
\end{figure}

We consider a subset of the data set available at \url{http://archive.ics.uci.edu/ml/datasets/Air+Quality} (\cite{de2008field, de2009co, de2012semi}).
The data set contains hourly averaged concentration measurements for different atmospheric pollutants obtained at a main road in an Italian city. Here we analyze measurements from June to September 2004, which result in 2928 observations.
The measurements for the pollutants were taken from two different sensors, standard (high-cost) sensors and  new low-cost (lc) sensors. We refer to a value measured with the standard (high-cost) sensor as a ground truth (gt) value. Ground truth values are available for CO (mg/$\mbox{m}^3$),  NOx (ppb) and NO2 ($\mu$g/$\mbox{m}^3$) and the aim is to predict these values. For each ground truth value we are given a corresponding value obtained from a low-cost sensor, resulting in six different pollution measurements for one time point. The measurements in July for the pollutant CO are visualized in Figure \ref{fig:data}. We see that the measurements of the ground truth sensor for CO are missing for several days, i.e. missing observations are present in this data set. The missing values per pollutant range from $4\%$ to $24\%$, whereas ground truth values have a higher portion of missing values.
In addition to the pollution measurements, hourly measurements of the temperature and of relative humidity are also available.

 In the following, the data containing the pollutant measurements is denoted by $y_{tj}, j=1, \ldots, 6, t=1, \ldots, T$, where $T=2928$ is the length of the time series. As before, $\mathcal {T}_j^{obs}$ is the set of time indices for which observed values are available for the $j$-th marginal time series. The measurements of relative humidity and temperature are denoted by $\text{TEMP}_{t}$ and $\text{RH}_{t}$, respectively for $t=1, \ldots, T$. 

\subsection{Marginal models}
\label{sec:dataana_marg}
We fit a generalized additive model (GAM) for each pollutant, where temperature, relative humidity, the hour at time $t$, $\text{H}_t\in \{0,\ldots, 23\}$, and the day at time $t$, $\text{D}_t \in \{0,\ldots, 6\}$  are used as covariates. We denote the covariates by $\boldsymbol{x_t} = (\text{TEMP}_{t}, \text{RH}_{t}, \text{H}_{t}, \text{D}_{t})$. As explained in Section \ref{marginal}, we allow for Box-Cox transformations (\cite{box1964analysis})
and assume that
\begin{equation}
BC(Y_{tj},\lambda_j) = f_j(\boldsymbol{{x}_{t}}) + \sigma_j\epsilon_{tj}
\end{equation}
with $\epsilon_{tj} \sim N(0,1)$ for $t =1, \ldots, T, j=1, \ldots, 6$ and $BC(Y_{tj},\lambda_j)$ as in \eqref{eq:boxcox}.

\begin{figure}[H]
\centerline{%
\includegraphics[trim={0 0cm 0 0},width=0.9\textwidth]{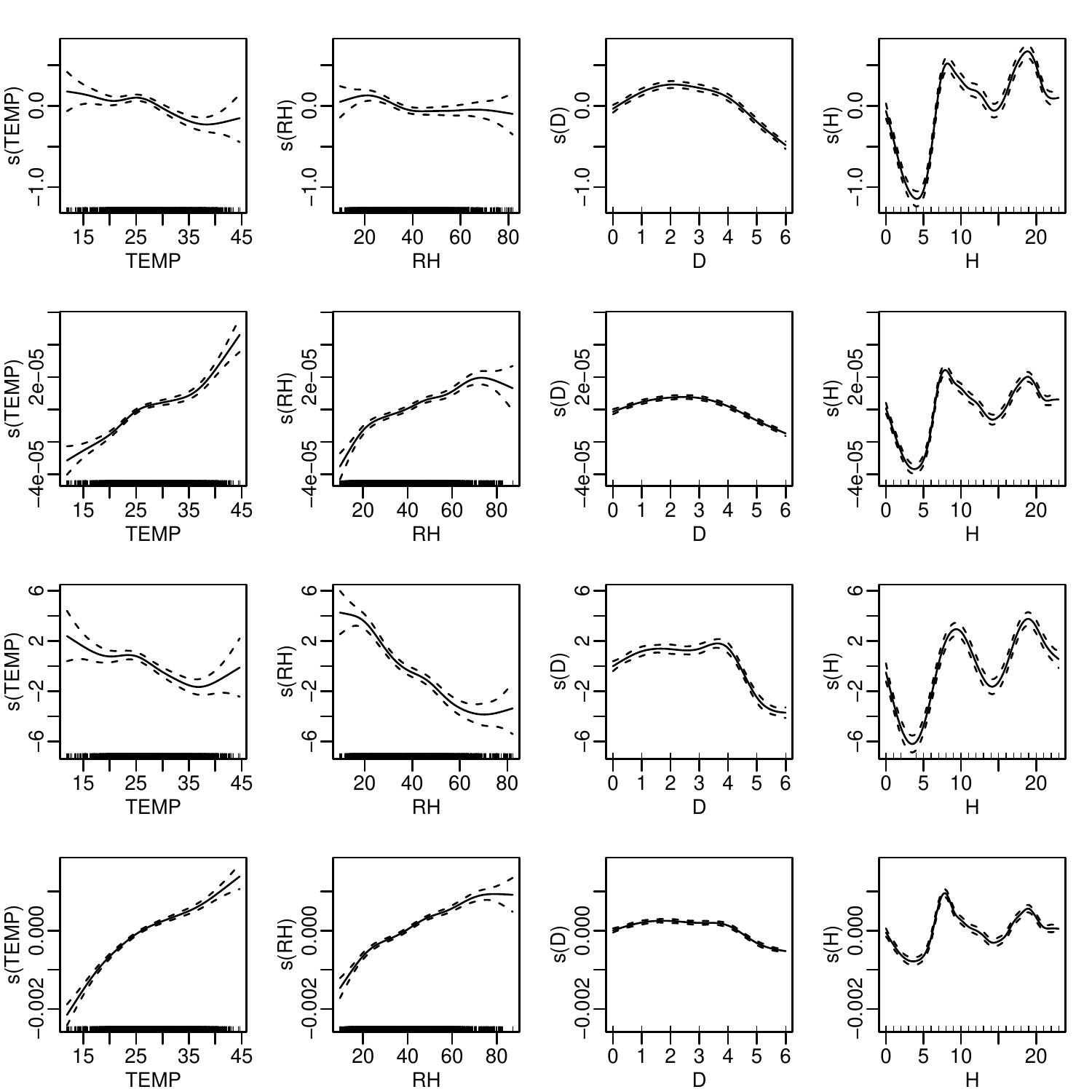}%
}%
\caption{Estimated smooth components of the GAMs for four Box-Cox transformed pollutants: CO(gt), CO(lc), NO2(gt), NO2(lc) (top to bottom row). Each GAM has four covariates, TEMP, RH, D and H.  The dashed lines represent a pointwise $95\%$ confidence band.}
\label{fig:gam}
\end{figure}

 For estimating the conditional mean function $f_j$ and $\sigma_j$ we assume that the errors $\epsilon_{tj}$ are independent. Later the dependence among the errors will be modeled with the proposed state space model. We estimate a GAM for different values of $\lambda_j$ and then choose the model which maximizes the likelihood for given data $y_{tj}, t \in \mathcal{T}_j^{obs}, j=1, \ldots, 6$. 
For each GAM we remove the corresponding  missing values and rely on the R package \texttt{mgcv} of \cite{wood2015package} for parameter estimation. We obtain estimates $\hat f_j$, $\hat \sigma_j$ and $\hat \lambda_j$ for $j=1, \ldots, 6$. From Table \ref{tab:lambdaest}, we see that the estimates for $\lambda_j$ deviate from 1, which indicates that the Box-Cox transformations are necessary. Figure \ref{fig:gam} shows the smooth components of the GAM for four different pollutants. We see for example a nonlinear effect of the Hour on the pollution measurement. The pollution is high at around 8 am and at around 6 pm, which may correspond to the hours with the highest traffic due to commuting workers.

\begin{table}[ht]
\centering
\begin{tabular}{rrrrrrr}
  \hline
 & CO(gt) & CO(lc) &  NOx(gt) & NOx(lc) & NO2(gt) & NO2(lc)  \\
  \hline
$\hat \lambda_j$ & 0.15 & -1.25 & 0.05 & 0.05 & 0.55 & -0.70  \\ 
   \hline
\end{tabular}
\caption{Estimates of $\lambda_1, \ldots, \lambda_6$ for the six GAMs fitted to the six pollution measurements.}
\label{tab:lambdaest}
\end{table}

\subsection{Dependence Model}
\label{sec:depmodel}
\begin{figure}[H]
\centerline{%
\includegraphics[trim={0 1cm 0 0},width=0.95\textwidth]{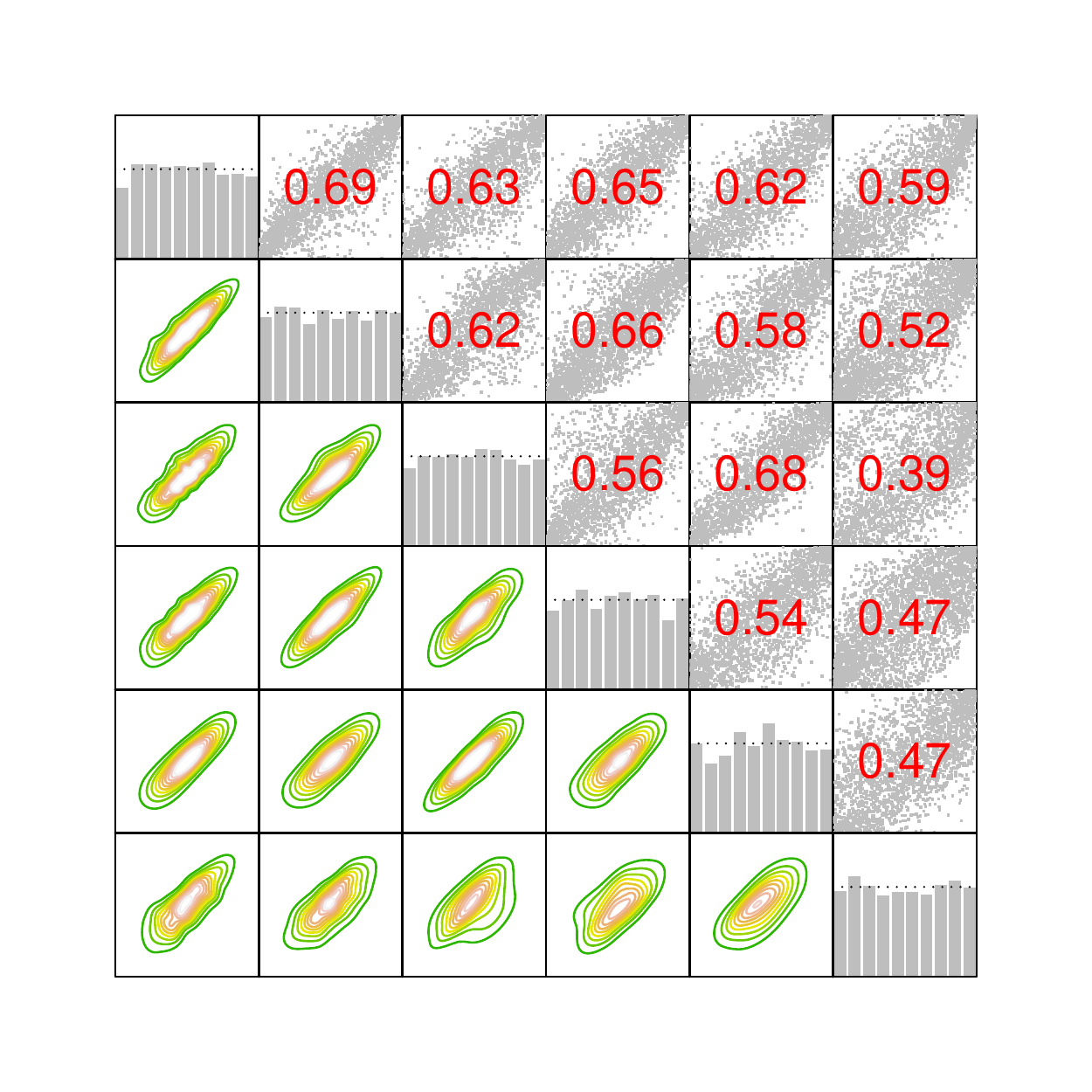}%
}%
\caption{The lower triangular part shows contour plots of all pairs of $(\hat z_{t1}, \ldots, \hat z_{t6}), t=1, \ldots, T$ ignoring serial dependence. The upper triangular part shows corresponding scatter plots of all pairs of $(\hat u_{t1}, \ldots, \hat u_{t6}), t=1, \ldots, T$ with $ \hat u_{tj}=\Phi(\hat z_{tj})$ and the empirical Kendall's $\tau$ for each pair. The diagonal shows the histograms of the univariate marginals. More precisely, the plot in the $i$-th row and $j$-th column shows the contour plot for the pair $(\hat z_{ti}, \hat z_{tj})$ if $i>j$, the scatter plot of $(\hat u_{ti}, \hat u_{tj})$ if $i<j$, or the histogram of $\hat u_{ti}$, if $i=j$. The variables are ordered as follows: 1: CO(gt), 2: CO(lc), 3: NOx(gt), 4: NOx(lc), 5: NO2(gt), 6: NO2(lc).}
\label{fig:contour1}
\end{figure}

\begin{figure}[H]
\centerline{%
\includegraphics[trim={0 6.25cm 0 0},width=0.9\textwidth]{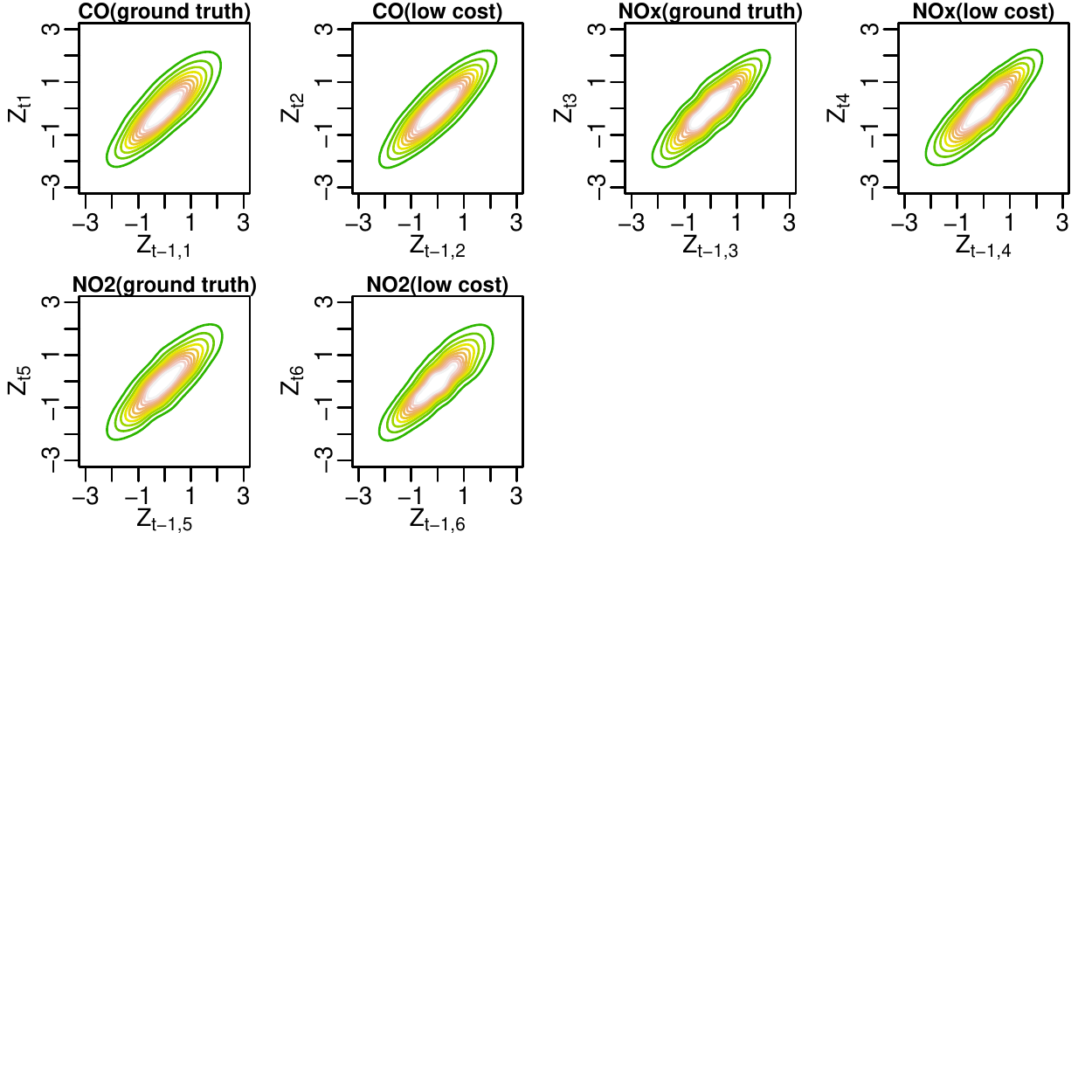}%
}%
\caption{Contour plots of pairs $(\hat{z}_{tj}, \hat{z}_{t-1j})_{t=2, \ldots, T}$ for $j=1, \ldots, 6$ ignoring serial dependence.}
\label{fig:contourtime1}
\end{figure}

Recall the standardized errors $Z_{tj}$, defined in \eqref{eq:gamerr}, as
\begin{equation*}
Z_{tj} = \frac{BC(Y_{tj},\lambda_j) - f_j(\boldsymbol{{x}_{t}})}{\sigma_j}
\end{equation*}
which are $N(0,1)$ distributed.
Pseudo observations of $Z_{tj}$ can be obtained from the estimates $\hat f_j$, $\hat \sigma_j$ and $\hat \lambda_j$ as
\begin{equation}
\hat z_{tj} = \frac{BC(y_{tj}, \hat \lambda_j) - \hat f_j(\boldsymbol{{x}_{t}})}{\hat \sigma_j}
\end{equation}
for $t=1, \ldots, T, j=1, \ldots 6$.
To visualize cross-sectional dependencies among the variables $Z_{tj}$ we examine bivariate contour plots for all pairs of $(\hat z_{t1}, \ldots, \hat z_{t6}), t = 1, \ldots, T$ in Figure \ref{fig:contour1}, ignoring serial dependence.  In addition, we examine contour plots of pairs  $(\hat z_{tj},\hat z_{t-1j}), t=2, \ldots, T$ for $j=1, \ldots, 6$ in Figure \ref{fig:contourtime1} to visualize temporal dependence. We observe temporal and cross-sectional dependence. Further, the dependence structures seem to be different from a Gaussian one since we observe asymmetries in the contour plots. For example, the contour plot in the bottom left corner of Figure \ref{fig:contour1} indicates stronger dependence in the upper right corner than in the bottom left corner. Therefore, a linear Gaussian state space model might not be appropriate here, but the proposed copula-based state space model can be a good candidate for this data.

\begin{figure}[H]
\centerline{%
\includegraphics[trim={0 3.5cm 0 0},width=0.85\textwidth] {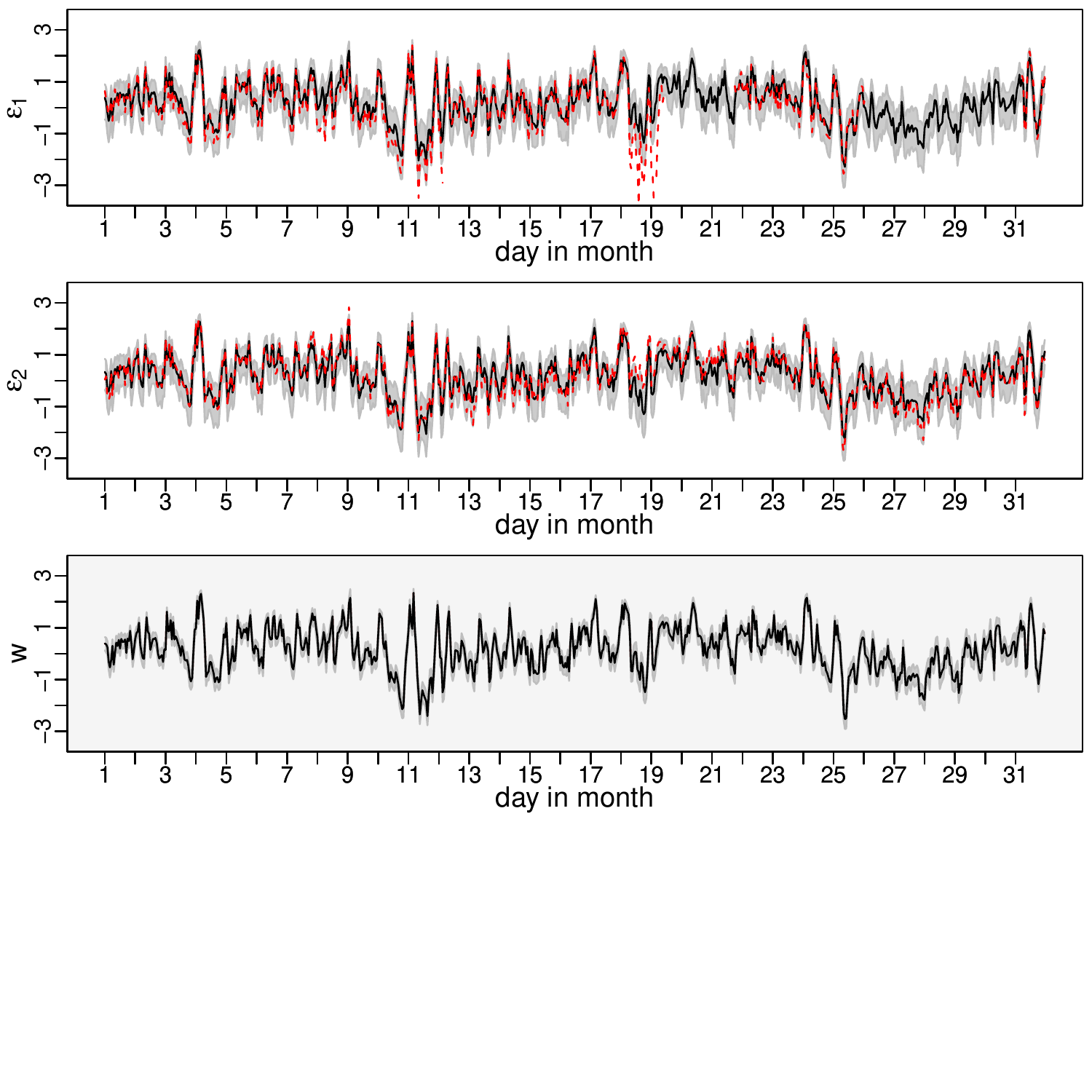}
}%
\caption{This plot is based on data for July. We show the estimated posterior mode of the predictive distribution of the standardized error $\epsilon_{tj}$ plotted against time $t$ for $j=1, 2$, corresponding to CO(gt) and CO(lc). Draws of the predictive distribution of the error are obtained as $\epsilon_{tj}^r = \Phi^{-1}(u_{tj}^r)$, where $u_{tj}^r$ is a sample from the predictive distribution on the copula scale. The observed standardized residual from the GAM is added in red (dashed). In addition, we show the estimated posterior mode of $w_t=\Phi^{-1}(v_t)$ plotted against $t$ in the third row. To all plots we add a $90\%$ credible region constructed from the estimated $5\%$ and $95\%$ posterior quantiles.}
\label{fig:U_sim1}
\end{figure}

Since our multivariate copula state space model operates on marginally uniform(0,1) distributed data, we obtain uniform(0,1) distributed random variables as follows
\begin{equation}
U_{tj} = \Phi(Z_{tj})
\end{equation}
with corresponding pseudo observations
\begin{equation}
 \hat u_{tj} = \Phi(\hat z_{tj})   
\end{equation}
for $t\in \mathcal{T}_j^{obs}, j=1, \ldots, 6$.
The proposed multivariate copula-based state space model is fitted to the data $\hat u_{tj}, t\in \mathcal{T}_j^{obs}, j=1, \ldots, 6$. Plots of the estimated posterior densities and trace plots are shown in the supplementary material. These plots indicate proper mixing of the Markov Chain. Table \ref{tab:fam_sel} shows the selected copula families corresponding to the estimated posterior modes of $m_{obs,j}$ or $m_{obs}$. We see that four Gaussian, one Student t and two Gumbel copulas were selected. In particular, our model features an asymmetric dependence structure, since the Gumbel copula is included.
Simulations of the in-sample period predictive distribution can be obtained as explained in Section \ref{sec:pdist_ins}. Transforming these simulations with the standard normal quantile function, we obtain predictive simulations for the standardized errors, i.e. we obtain draws from the predictive distribution of the error as
\begin{equation}
\epsilon_{tj}^r = \Phi^{-1}(u_{tj}^r),
\end{equation}
for $r=1, \ldots, 3000$, where $u_{tj}^r$ is a draw from the in-sample predictive distribution on the copula scale (see Section \ref{sec:bayesinf}).
These simulations are compared to the observed standardized residual of the GAM ($\hat z_{tj}$) to assess how well our model fits the data. In particular, we want to asses if a single factor structure is appropriate or if it is too restrictive. 
According to Figure \ref{fig:U_sim1}, the model seems to be appropriate. The single factor structure is able to capture the time dynamics of the residuals. The ground truth values for CO are missing from day 26 to day 30. We see that within this period the time dynamic is learned from other series where data is available within this period.     While Figure \ref{fig:U_sim1} shows plots for two pollutants in July, plots for different pollutants in different months looked similar.

\begin{table}[H]
\centering
\begin{tabular}{rlllllllll}
  \hline
 & $\hat m_{obs,1}$ & $\hat m_{obs,2}$ & $\hat m_{obs,3}$ & $\hat m_{obs,4}$ & $\hat m_{obs,5}$ & $\hat m_{obs,6}$ & $\hat m_{lat}$ \\ 
  \hline
Copula family & Gu & Gu & Ga & S & Ga & Ga & Ga \\ 
   \hline
\end{tabular}
\caption{The marginal posterior mode estimates of the copula family indicators $\boldsymbol m_{obs},m_{lat}$. (Ga: Gaussian, S: Student t(df=4), C: Clayton, Gu: Gumbel).}
\label{tab:fam_sel}
\end{table}

\subsection{Predictions}
\label{sec:pred}
We evaluate the proposed model's ability to predict the ground truth values. Therefore we compare the copula state space model to a Gaussian state space model and to Bayesian additive regression trees (\cite{chipman2010bart}), as a representative for a popular machine learning algorithm. Compared to other machine learning techniques, Bayesian additive regression trees
have the advantage that a predictive distribution is obtained instead of a single point estimate. Therefore we can compare models with respect to their forecast distribution, for which we utilize the continuous ranked probability score (\cite{gneiting2007strictly}). The continuous ranked probability score (CRPS) for an observed value $y \in \mathbb{R}$ and a univariate forecast CDF $F$ is defined as
\begin{equation}
CRPS =    \int_{\mathbb{R}} (F(z) - 1_{y\leq z})^2 dz.
\end{equation}

For each of the ground truth values we remove the observations in the last month of the data set and treat them as missing values, which yields the training set. Based on the training set we proceed similarly to what we described above, i.e. we first estimate the GAMs, and then estimate the state space model on the copula scale. Here two state space models are estimated: the copula state space model where the family set $\mathcal{M}$ is chosen as in Section \ref{sec:depmodel} and the Gaussian state space model where we restrict the family set to $\mathcal{M}=\{\text{Gaussian}\}$.
 For each of the two state space models we obtain $2000$ simulations from the in-sample predictive distribution $u_{tj}^r$, $r=1, \ldots, 2000$, whereas our MCMC approach of Section \ref{sec:bayesinf} is run for 3000 iterations and the first 1000 draws are discarded for burn-in. Here $t$ is a timepoint which is among the newly selected missing values for the ground truth value that corresponds to margin $j$. Based on these simulations we obtain simulations from the predictive distribution of the Box-Cox transformed response as follows
\begin{equation}
(y_{tj}^{bc})^r=\hat f_j(\boldsymbol {x_t}) + \hat \sigma_j \Phi^{-1}(u_{tj}^r)
\end{equation}
for $r=1, \ldots, 2000.$

Since Bayesian additive regression trees rely on the normal distribution, we expect that Box-Cox transformations might also improve the fit for this model. We assume that
\begin{equation}
BC(Y_{tj}, \lambda_j) = g_j(\boldsymbol{x_{tj}^{BART}}) +  \sigma_j \epsilon_{tj},
\end{equation}
where $\boldsymbol{x_{tj}^{BART}}$ are the covariates, $g_j(\cdot)$ is a sum of regression trees and $\epsilon_{tj} \sim N(0,1)$.
In addition to the covariates used for the GAM model, all pollutant measurements except the one corresponding to margin $j$ are included in the covariate vector $\boldsymbol {x_{tj}^{BART}}$. For $ \lambda_j$ we use the same value as for the previously fitted GAM. We have seen that this transformation improves the performance of the Bayesian additive regression trees. \cite{mcculloch2018bart} implement a MCMC sampler in the \texttt{R} package \texttt{BART} which we use to obtain draws $g_j^r, \sigma_j^r$, $r=1, \ldots, 10000$ from the corresponding posterior distribution. We discard the first 5000 of these draws and then 5000 simulations of the predictive distribution of the response are obtained as
\begin{equation}
(y_{tj}^{bc})^r \sim N(g_j^r(\boldsymbol{x^{BART}_{tj}}), (\sigma_j^r)^2)    
\end{equation}
for $r=1, \ldots, 5000$.

Based on the simulations $(y_{tj}^{bc})^r, r=1, \ldots, 5000$ we calculate the empirical CDF and use this to approximate the CRPS (this is implemented in the \texttt{R} package \texttt{scoringRules} of \cite{jordan2017evaluating}) for the different time points and sum them up to obtain the cumulative CRPS. For each of the three methods, we obtain a cumulative CRPS for each the three ground truth indices.
In addition, we consider reduced bivariate data sets, where each data set consists of the ground truth value of a pollutant, the corresponding low-cost value and the covariates as in Section \ref{sec:dataana_marg}. This yields three reduced data sets, each associated with one of the three pollutants. For each of the reduced data sets we proceed as above, i.e. we first remove ground truth observations in the last month, fit the three different models and calculate the CRPS values.

We refer to the models fitted to the reduced data as bivariate state space models and reduced Bayesian additive regression trees. The models estimated with the full data are referred to as joint models. 
We want to investigate how the bivariate state space models compare to the six-dimensional ones.
The cumulative CRPS values are shown in Table \ref{tab:crps}. 
For the pollutant NOx, the state space approach seems not to be the best choice. We have seen (see supplementary material) that for this pollutant, the dependence between the ground truth and the low-cost values varies more over time than for the other pollutants. Relaxing the assumption of a time-constant Kendall's $\tau$, might improve the predictive accuracy for this pollutant. This model extension is subject to future research.
Overall, the copula state space model is the best performing model within this comparison, since it outperforms the Gaussian state space model and the Bayesian additive regression trees in two out of three cases. 

\begin{table}[H]
\centering
\begin{tabular}{l|rrr}
  \hline
 & CO & NOx & NO2  \\ 
  \hline
joint copula state space model  & \textbf{74.27} & 594.50 & \textbf{569.03}   \\
bivariate copula state space model  & 84.92 & 559.22 & 845.95   \\ 
joint Gaussian state space model & 76.91 & 594.30 & 570.55   \\ 
bivariate Gaussian state space model  & 87.90 & 559.18 & 844.64   \\ 
joint Bayesian additive regression trees & 183.49 & \textbf{379.31} & 1330.90   \\ 
reduced Bayesian additive regression trees & 89.39 & 520.93 & 1095.40   \\ 
   \hline
\end{tabular}
\caption{Cumulative CRPS for the three ground truth values (CO, NOx, NO2) obtained from six different models: joint/bivariate copula state space model, joint/bivariate Gaussian state space model, joint/reduced Bayesian additive regression trees. The best, i.e. the lowest, cumulative CRPS value in marked in bold.}
\label{tab:crps}
\end{table}

\section{Concluding Remarks}\label{Conclusions}
We proposed a multivariate nonlinear non-Gaussian copula-based state space model. The model is very flexible: the observation and the state equation are specified with copulas and the model can be combined with different marginal distributions. We illustrated the model with air pollution measurements data and have shown that the novel copula state space model outperforms a linear Gaussian state space model and Bayesian additive regression trees. 
As we have seen in Section \ref{sec:pred}, the assumption of a time-constant dependence structure might not always be appropriate. A first extension of the model could allow for dynamic dependence parameters. For this, ideas of the dynamic bivariate copula model of \cite{almeida2012efficient} might be used. Another area of future research is the extension to multiple factors.

\bibliographystyle{spbasic}    
\bibliography{References}{}

\newpage
\noindent
{\Huge\bf Supplementary material}

\setcounter{section}{0}

\section{Additional Material for Section 2.2}
\label{sec:app_lingauss}
\vspace*{1cm}
The covariance matrix $\Sigma$ of the joint distribution 
$$
(Z_{11}, \ldots, Z_{d1}, W_1; Z_{12}, \ldots, Z_{d2}, W_2; \ldots, Z_{1T}, \ldots, Z_{dT}, W_T) \sim N_{(d+1)T} (\textbf{0}, \Sigma)
$$
takes the form
$$
\Sigma = \begin{pmatrix}
    A       & &   \rho_{lat}(A + B)   & &  \rho_{lat}^2(A + B)   & &  \ldots & & \rho_{lat}^{T-1}(A + B) \\ \\
    \rho_{lat}(A + B)      & & A   &&   \rho_{lat}(A + B)   & &  \ldots &&  \rho_{lat}^{T-2}(A + B)  \\ \\
    \rho_{lat}^2(A + B)   &&   \rho_{lat}(A + B)  & &   A  & &  \ldots &&  \rho_{lat}^{j-2}(A + B)   \\ \\
    \rho_{lat}^3(A + B)  &&   \rho_{lat}^2(A + B)    &&   \rho_{lat}(A + B)   &&   \ddots && \vdots \\ \\
    \vdots  &&  \vdots && \vdots && \vdots && \vdots \\ \\
    \rho_{lat}^{T-1}(A + B)  && \rho_{lat}^{T-2}(A + B) && \rho_{lat}^{T-3}(A + B) && \ldots && A 
\end{pmatrix}
$$

\vspace*{1cm}

where the matrices $A$ and $B$ take the following forms
\vspace*{1cm}

$$
A = \begin{pmatrix}
    1       &    \rho_{obs,1} \rho_{obs,2}   &   \rho_{obs,1} \rho_{obs,3}   & \ldots & \rho_{obs,1} \rho_{obs,d} &  \rho_{obs,1}  \\
    \rho_{obs,1} \rho_{obs,2}      & 1   &   \rho_{obs,2} \rho_{obs,3} & \ldots & \rho_{obs,2} \rho_{obs,d} &  \rho_{obs,2}  \\
    \rho_{obs,1} \rho_{obs,3}   &   \rho_{obs,2} \rho_{obs,3}   &   1 & \ldots  & \rho_{obs,3} \rho_{obs,d} &  \rho_{obs,3}  \\
    \vdots & \vdots & \vdots & \ddots & \vdots & \vdots \\
    \rho_{obs,1} \rho_{obs,d}   &   \rho_{obs,2} \rho_{obs,d} & \rho_{obs,3} \rho_{obs,d}   & \ldots  &   1 &  \rho_{obs,d}  \\
    \rho_{obs,1}  &   \rho_{obs,2}    &   \rho_{obs,3} & \ldots &   \rho_{obs,d} &   1 \\
\end{pmatrix}
$$
\vspace*{1cm}

$$
B = \begin{pmatrix}
    \rho_{obs,1}^2 - 1    &  0   &  0  & \ldots & 0  &   0\\
    0      & \rho_{obs,2}^2 - 1  & 0  & \ldots & 0 &    0 \\
    0   &   0   &   \rho_{obs,3}^2 - 1  & \ldots &  0 &  0 \\
    \vdots & \vdots & \vdots & \ddots & \vdots & \vdots \\
    0  &   0    &   0  & \ldots  & \rho_{obs,d}^2 - 1 & 0 \\
    0  &   0    &   0 & \ldots & 0 & 0
\end{pmatrix}.
$$

\section{Additional Material for Section 4.3}
\label{sec:app_depmodel}
\begin{figure}[H]
\centerline{%
\includegraphics[trim={0 0cm 0 0},width=0.8\textwidth] {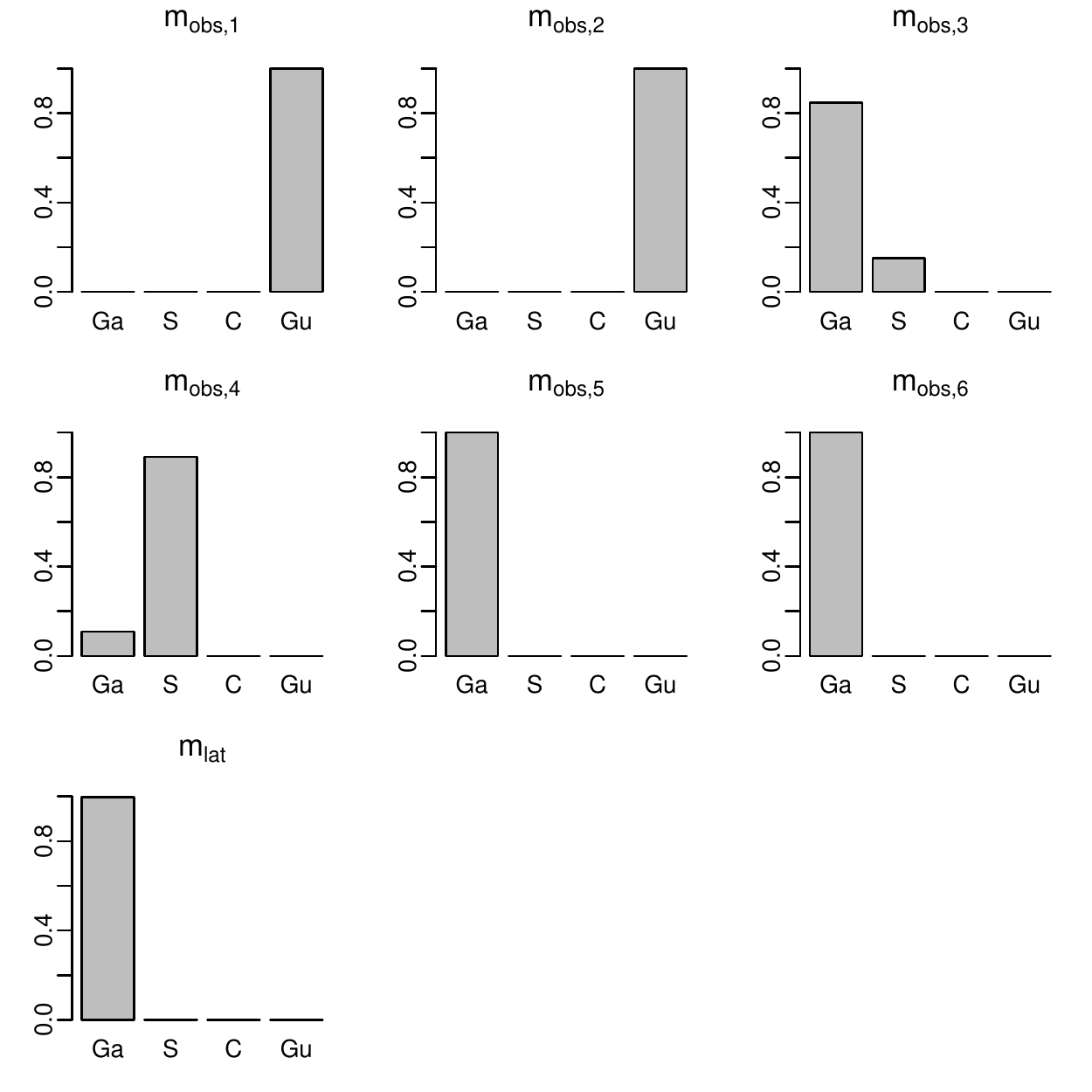}
}%
\caption{Estimated posterior distribution of the copula family indicators $m_{obs1}, \ldots, m_{obs,6}, m_{lat}$ obtained from 2000 iterations after a burn-in of 1000.}
\label{fig:post_dist_fam}
\end{figure}

\begin{figure}[H]
\centerline{%
\includegraphics[trim={0 6.5cm 0 0},width=0.75\textwidth] {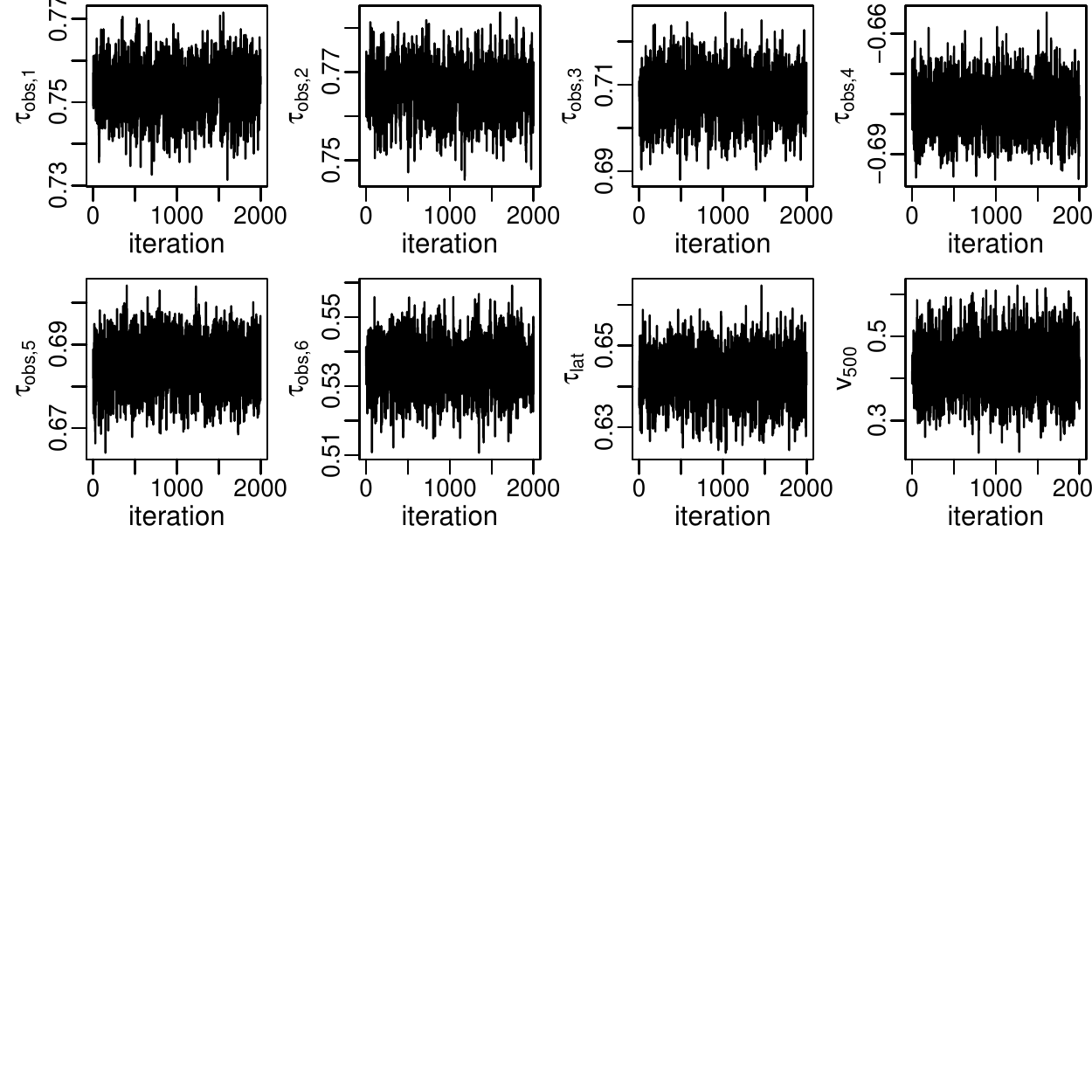}
}%
\caption{Trace plots of 2000 draws after a burn-in of 1000 for selected parameters of the copula state space model. The variables are ordered as follows: 1: CO(gt), 2: CO(lc), 3: NOx(gt), 4: NOx(lc), 5: NO2(gt), 6: NO2(lc).}
\label{fig:trace}
\end{figure}

\begin{figure}[H]
\centerline{%
\includegraphics[trim={0 6.5cm 0 0},width=0.75\textwidth] {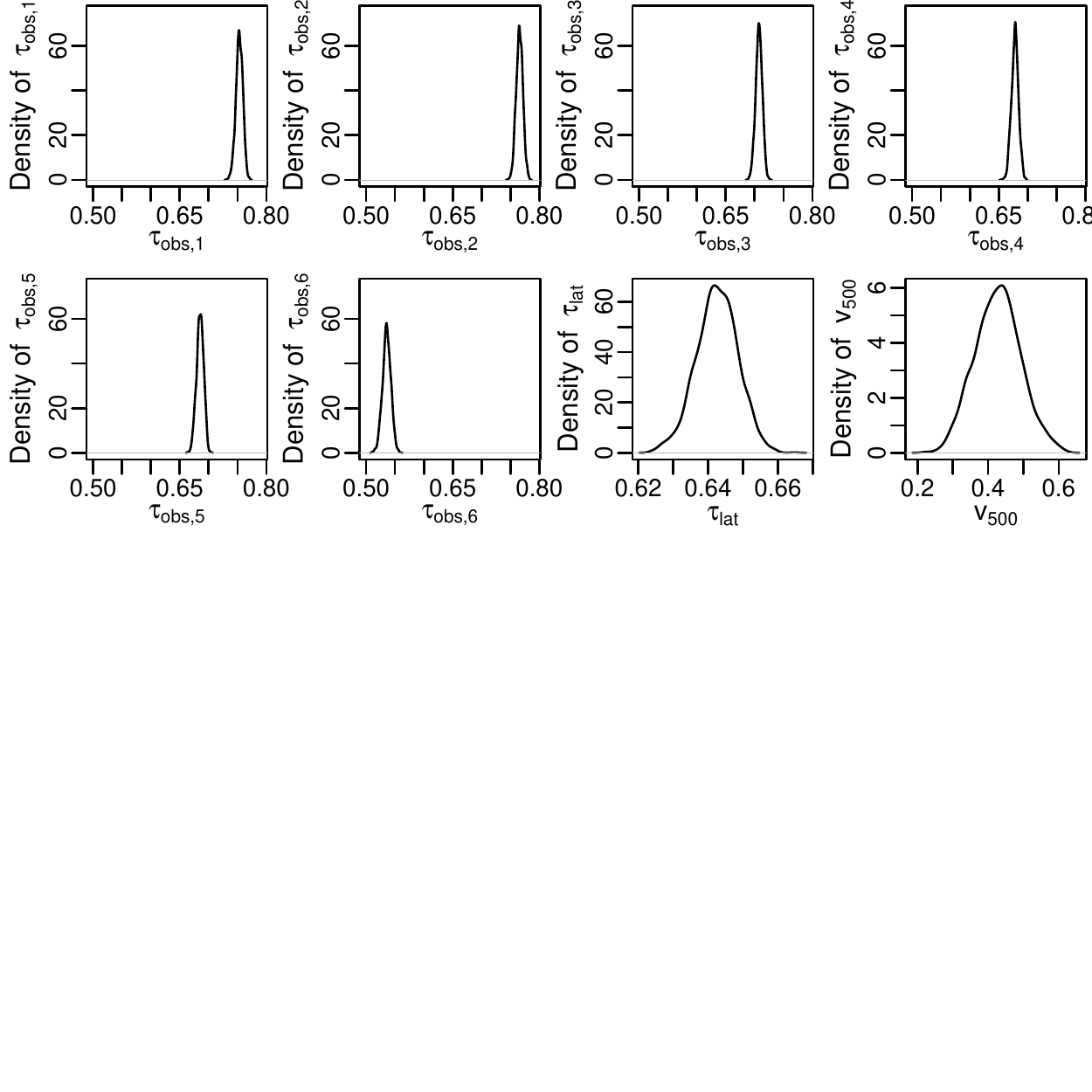}
}%
\caption{Estimated posterior density for selected parameters of the copula state space model. The posterior density is estimated as the kernel density estimate based on 2000 draws after a burn-in of 1000. For better comparability we multiplied the draws of $\tau_{obs,4}$ by $-1$. The variables are ordered as follows: 1: CO(gt), 2: CO(lc), 3: NOx(gt), 4: NOx(lc), 5: NO2(gt), 6: NO2(lc).}
\label{fig:dens}
\end{figure}

\section{Additional Material for Section 4.4}
\label{sec:app_pred}

\begin{figure}[H]
\centerline{%
\includegraphics[trim={0 0cm 0 0},width=0.8\textwidth] {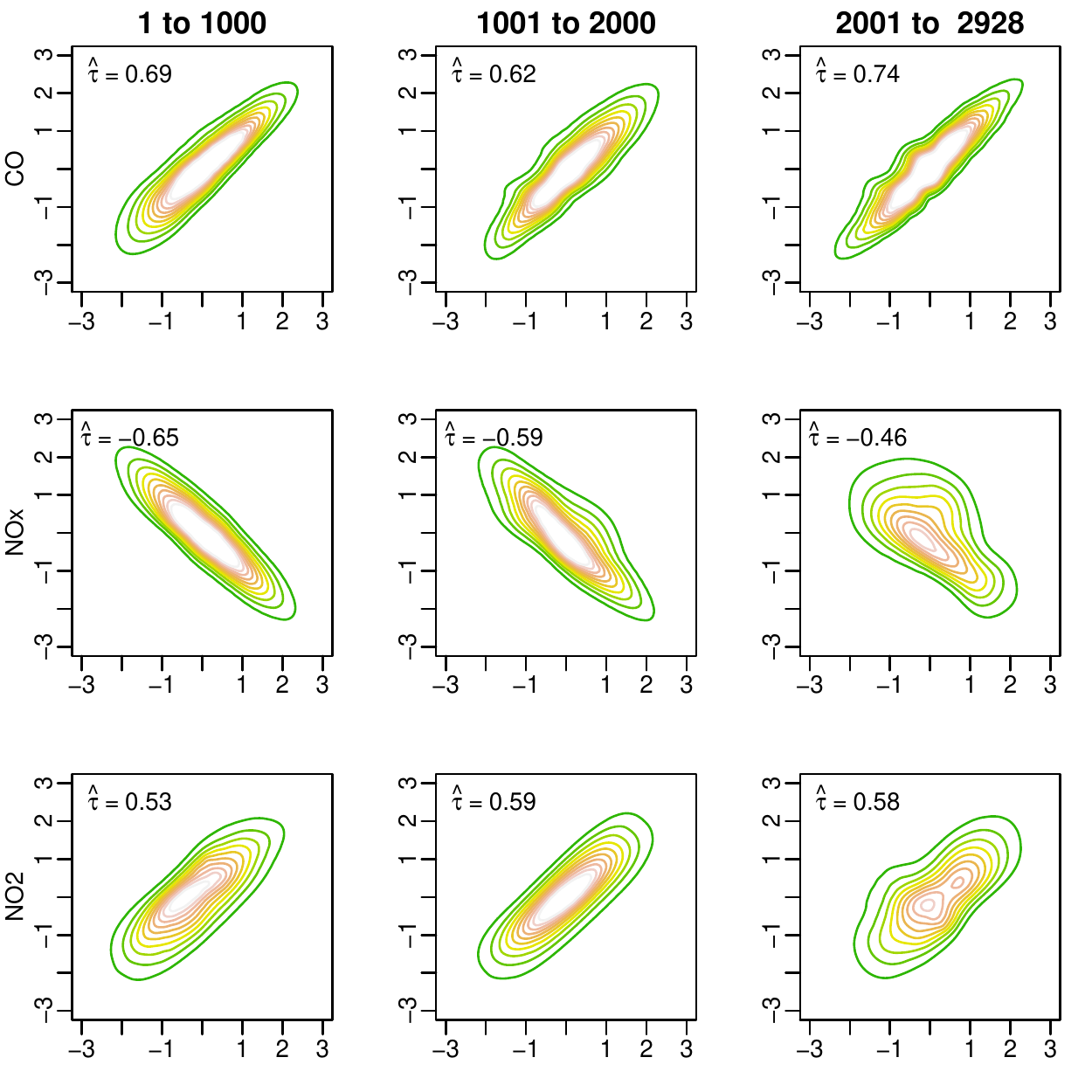}
}%
\caption{Contour plots for pairs ($\hat z_{tj}, \hat z_{tj'})_{t \in P_i}$, $i=1,2,3$ where $j$ corresponds to a ground truth and $j'$ to the corresponding low-cost value within a time period $P_i$ ($P_1: 1, \ldots, 1000$, $P_2: 1001, \ldots, 2000$, $P_3: 2001, \ldots, 2928$). For example the top row shows contour plots for the (CO(gt), CO(lc)) pair for the three different time periods. In the top left corner we added the corresponding empirical Kendall's $\tau$,  based on the data $(\hat z_{tj}, \hat z_{tj'})_{t \in P_i}$.
}
\label{fig:app_cont}
\end{figure}

\end{document}